\documentclass[10pt,twocolumn,twoside]{IEEEtran}
\usepackage{times}
\usepackage[final]{graphicx}
\usepackage[reqno]{amsmath}
\usepackage{amsfonts}

\usepackage{times,amsmath,epsfig}
\usepackage{latexsym,amssymb}
\usepackage[center]{caption}
\usepackage{graphicx}         
\usepackage{color}            

\usepackage{marginnote}
\usepackage{multirow}
\usepackage{subfigure}

\newcommand{\Prob}{\textrm{Pr}}
\newcommand{\beq}{\begin{equation}}
\newcommand{\enq}{\end{equation}}
\newcommand{\beqa}{\begin{eqnarray}}
\newcommand{\enqa}{\end{eqnarray}}
\newcommand{\beqn}{\begin{eqnarray*}}
\newcommand{\enqn}{\end{eqnarray*}}

\newcommand{\no}{\nonumber}

\newtheorem{theorem}{Theorem}

\newtheorem{lemma}[theorem]{Lemma}

\newcommand{\qed}{\hfill $\Box$}

\allowdisplaybreaks[4]

\def\draftmode{}
\ifx\draftmode\undefined
\newcommand{\comment}[1]{}
\else
\newcommand{\comment}[1]{ \marginnote{$\Longleftarrow$}{\bf $<$#1$>$} }
\fi

\ifCLASSINFOpdf
\else
\fi
\begin{document}
%
\title{Keys through ARQ: Theory and Practice}
%
%
%

\author{
        Yara~Abdallah*,~\IEEEmembership{Student~Member,~IEEE,}
        Mohamed~Abdel~Latif,~\IEEEmembership{Student~Member,~IEEE,}
        Moustafa~Youssef,~\IEEEmembership{Senior~Member,~IEEE,}
        Ahmed~Sultan,~\IEEEmembership{Member,~IEEE,}
        and~Hesham~El~Gamal,~\IEEEmembership{Fellow,~IEEE}
\thanks{This work is funded in part by NSF, QNRF, USAID and the Egyptian Science and Technology Development Fund (STDF) under the US-Egypt Joint Research Grants Program. The material in this paper was presented in part at the
Communication theory symposium, International Conference of Communications, Dresden, Germany, June, 2009,
the IEEE Information Theory Workshop, Taormina, Sicily, Italy, October, 2009,
and the IEEE Information Theory Workshop, Cairo, Egypt, January, 2010.}
\thanks{Y. Abdallah, M. Youssef and A. Sultan are with the Wireless Intelligent Networks Center (WINC),
Nile University, Cairo, Egypt (e-mail: {yara.abdallah, asultan, mayoussef}~@nileu.edu.eg).}
\thanks{M. Abdel Latif was with the Wireless Intelligent Networks Center (WINC), Nile University, Cairo, Egypt
and is now with the Department of Electrical and Computer Engineering, University
of California, Irvine, CA, 92717 USA (e-mail: mohamed.abdellatif@uci.edu).}
\thanks{H. El Gamal is with the Department of Electrical and Computer Engineering,
Ohio State University, Columbus, OH, 43210 USA (e-mail: helgamal@ece.osu.edu).}
}

\maketitle

\begin{abstract}

This paper develops a novel framework for sharing secret keys using the \textbf{A}utomatic \textbf{R}epeat re\textbf{Q}uest (\textbf{ARQ}) protocol. We first characterize the underlying information theoretic limits, under different assumptions on the channel spatial and temporal correlation function. Our analysis reveals a novel role of ``dumb antennas" in overcoming the negative impact of spatial correlation on the achievable secrecy rates. We further develop an adaptive rate allocation policy, which achieves higher secrecy rates in temporally correlated channels, and explicit constructions for ARQ secrecy coding that enjoy low implementation complexity. Building on this theoretical foundation, we propose a unified framework for ARQ-based secrecy in Wi-Fi networks. By exploiting the existing ARQ mechanism in the IEEE 802.11 standard, we develop security overlays that offer strong security guarantees at the expense of only minor modifications in the medium access layer. Our numerical results establish the achievability of non-zero secrecy rates even when the eavesdropper channel is less noisy, on the average, than the legitimate channel, while our linux-based prototype demonstrates the efficiency of our ARQ overlays in mitigating all known, passive and active, Wi-Fi attacks at the expense of a minimal increase in the link setup time and a small loss in throughput.

\end{abstract}


%
\IEEEpeerreviewmaketitle

\section{Introduction}

The recent flurry of interest on wireless physical layer secrecy is inspired by Wyner's pioneering work on the wiretap channel~\cite{Wyner1975} which establishes the achievability of perfectly secure communication by hiding the message in the additional noise level seen by the eavesdropper. More recently, the effect of fading on the secrecy capacity was studied in which it was shown that, by appropriately distributing the message across different fading realizations, the multi-user diversity gain can be harnessed to enhance the secrecy capacity, e.g.~\cite{Gopala2008,Tang2007a}. Independent and parallel to our work, the authors of~\cite{Tang2007,Xiao2007,Xiao2010} considered using the well-known Hybrid ARQ protocol to facilitate the exchange of secure messages over fading channels. One innovative aspect of our framework, compared to~\cite{Tang2007}, is the distribution of key bits over an asymptotically large number of ARQ epochs. This approach allows for overcoming the secrecy outage phenomenon observed in~\cite{Tang2007} at the expense of increased delay. Contrary to~\cite{Xiao2010}, we build an information theoretic foundation for key sharing through ARQ which inspires low complexity implementation of practical coding schemes and reveals a novel role of dumb antennas in overcoming  the negative impact of spatial correlation, between the legitimate and eavesdropper channels, on the achievable key rate. Moreover, we propose a new greedy rate adaptation algorithm that is capable of transforming the temporal correlation in the legitimate channel into additional gains in the secrecy rate.

Building on our information theoretic foundation, we develop a unified ARQ security framework for Wi-Fi networks (ARQ-seCuRity fOr Wireless Networks: ARQ-CROWN); another distinguishing feature of our work as compared with~\cite{Tang2007,Xiao2007,Xiao2010}. This framework is used to construct \emph{security overlays} which provide information theoretic confidentiality guarantees to complement the underlying Wi-Fi security protocols. More specifically, careful analysis of the state of the art attacks on these protocols (e.g.,~\cite{Tews2008,Ohigashi,Khan2008}) reveals that they depend critically on the availability of certain security parameters as plaintext in the transmitted packets. By judiciously using the available ARQ mechanism in the IEEE 802.11 standard, our overlays transform those security parameters into a secret key that is shared only by the legitimate nodes. Remarkably, this goal is achieved through only minor modifications in the MAC layer that treat all protocols uniformly, and hence, does not entail additional network management tasks. The experimental results, obtained from our Madwifi driver prototype, demonstrate the ability of ARQ-CROWN to defend against all known eavesdropping attacks (whether active or passive), at the expense of a minor loss in throughput and a small increase in link setup time. This, to the best of our knowledge, the first attempt to demonstrate the utility of information theoretic security concepts in practice.

The remainder of this paper is organized as follows. We develop our information theoretic foundation in Section~\ref{sec_theoretic_foundation}. The design of our ARQ secrecy framework for Wi-Fi networks is presented in Section~\ref{sec_wifi}. Our numerical and experimental results are given in Section~\ref{sec_results}. Section~\ref{sec_conclusions} offers some concluding remarks whereas the proofs are collected in the appendices to enhance the flow of the paper.

\section{Information Theoretic Foundation}\label{sec_theoretic_foundation}

\subsection{System Model and Notations}\label{subsec_system_model}
Our model assumes one transmitter (Alice), one legitimate receiver (Bob), and one passive eavesdropper (Eve). We adopt a block fading model in which each channel is assumed to be fixed over one coherence interval and changes from one interval to the next. In order to obtain rigorous information theoretic results, we consider the scenario of asymptotically large coherence intervals and allow for sharing the secret key across an asymptotically large number of those intervals. The finite delay case will be considered in Section~\ref{subsec_explicit_coding}. In any particular interval, the signals received by Bob and Eve are respectively given by,
\begin{eqnarray*}
    y(i,j)&=&g_b(i)\, x(i,j)+w_b(i,j),\\
    z(i,j)&=&g_e(i)\, x(i,j)+w_e(i,j),
\end{eqnarray*}
\noindent where $x(i,j)$ is the $\it {j}^{th}$ transmitted symbol in the $\it {i}^{th}$ block, $y(i,j)$ is the $\it {j}^{th}$ received symbol by Bob in the $\it {i}^{th}$ block, $z(i,j)$ is the $\it {j}^{th}$ received symbol by Eve in the $\it {i}^{th}$ block, $g_{b}(i)$ and $g_{e}(i)$ are the complex block channel gains from Alice to Bob and Eve, respectively. The channel gains can also be written as $g_{b}(i) = \sqrt{h_{b}(i)}\exp(j\theta_b(i))$, and, $g_{e}(i) = \sqrt{h_e(i)}\exp(j\theta_e(i))$, where $\theta_b(i)$ and $\theta_e(i)$, the phase shifts at Bob and Eve respectively, are assumed to be independent in \textbf{all} considered scenarios. Moreover, $w_{b}(i,j)$ and $w_{e}(i,j)$ are the zero-mean, unit variance white complex Gaussian noise coefficients at Bob and Eve, respectively. We do not assume any prior knowledge about the
channel state information at Alice. Bob, however, is assumed to know $g_b(i)$ and Eve is assumed to know both $g_b(i)$ and $g_e(i)$ {\em a-priori}. We impose the following short-term average power constraint
\begin{eqnarray*}
    {\mathbb E} \left(|x(i,j)|^2\right)\leq \bar{P}.
\end{eqnarray*}
Our model only allows for one bit of ARQ feedback from Bob to Alice. Each ARQ epoch is assumed to be contained in one coherence interval (i.e., fixed channel gains) and that different epochs correspond to different coherence intervals. The transmitted packets are assumed to carry a perfect error detection mechanism allowing Bob (and Eve) to determine whether the packet has been received correctly or not. Bob sends back to Alice an ACK/NACK bit, through a public feedback channel which is only accessible by Bob but Monitored by Eve. To minimize Bob's receiver complexity, we adopt the memoryless decoding assumption implying that frames received in error are discarded and not used to aid in future decoding attempts. Finally, Eve is assumed to be passive (i.e., can not transmit); an assumption which can be justified in several practical settings. We will argue in Section~\ref{sec_wifi}, however, that our approach can mitigate all known active attacks on Wi-Fi networks as well.

In our setup, Alice wishes to share a secret key $W \in {\mathcal W}=\{1,2,\cdots ,M\}$ with Bob. To transmit this key, Alice and Bob use an $(M,m)$ code consisting of : 1) a stochastic encoder $f_m(.)$ at Alice that maps the key $w$ to a codeword $x^m \in {\mathcal X}^{m}$, 2) a decoding function $\phi$: ${\mathcal Y}^{m}\rightarrow {\mathcal W}$
which is used by Bob to recover the key. The codeword is partitioned into $a$ blocks, each one corresponds to one ARQ-epoch and contains $n_1$ symbols where $m = a\,n_1$. Unless otherwise stated, we focus on the asymptotic scenario where
$a\rightarrow\infty$ and $n_1\rightarrow\infty$. Alice starts with a random selection of the first block of $n_1$ symbols. Upon reception, Bob attempts to decode this block. If successful, it sends an ACK bit to Alice who moves ahead and makes a random choice of the second $n_1$ and sends it to Bob. Here, Alice must make sure that the concatenation of the two blocks belong to a valid codeword. As shown in the sequel, this constraint is easily satisfied. If an error was detected, then Bob sends a NACK bit to Alice; in which case both Alice and Bob will discard this block. Alice will then {\bf replace} the first block of $n_1$ symbols with another randomly chosen block and transmits it. The process then repeats until Alice and Bob agree on a sequence of $a$ blocks, each of length $n_1$ symbols, corresponding to the key. It is interesting to note that this strategy {\bf does not include any retransmissions}. The optimality of this approach, as proved in our main results, hinges on this property which minimizes the {\bf information leakage} to Eve.

The code construction must allow for reliable decoding at Bob while hiding the key from Eve. It is clear that the proposed protocol exploits the error detection mechanism to make sure that both Alice and Bob agree on the key (i.e., ensures reliable decoding). What remains is the secrecy requirement which is measured by the equivocation rate $R_e$ defined as the entropy rate of the transmitted key conditioned on the intercepted ACKs or NACKs and the channel outputs
at Eve, i.e.,
\begin{equation*}
    R_e ~\overset{\Delta}{=}~ \frac{1}{n} H(W|Z^n,K^b,G_b^b,G_e^b) ~,
\end{equation*}
\noindent where $n$ is the number of symbols transmitted to exchange the key (including the symbols in the discarded blocks due to decoding errors), $b=a\frac{n}{m}$, $K^b = \{ K(1), \cdots, K(b)\}$ denotes sequence of ACK/NACK bits, $G_b^b$ and $G_e^b$ are the sequences of channel coefficients seen by Bob and Eve in the $b$ blocks, and $Z^n = \{ Z(1), \cdots, Z(n)\}$ denotes Eve's channel outputs in the $n$ symbol intervals. We limit our attention to the \textbf{perfect secrecy} scenario, which requires the equivocation rate $R_e$ to be arbitrarily close to the key rate. The secrecy rate $R_s$ is said to be achievable if for any $\epsilon>0$, there exists a sequence of codes $(2^{nR_s},m)$ such that for any $m\geq m(\epsilon)$, we have
$R_e = \frac{1}{n} H(W|Z^n,K^b,G_b^b,G_e^b) ~\geq~ R_{s}-\epsilon$, and the {\bf key rate} for a given input distribution is defined as the maximum achievable perfect secrecy rate with this distribution.

%
%

\subsection{Main Result}\label{subsec_main_result}

Our main result is derived for the scenario where the feedback channel is error free and $h_e, h_b$ vary \textbf{independently} from one block to another according to a joint distribution $f\left(h_b,h_e\right)$. We will consider the effect of spatial and temporal correlation in Section~\ref{subsec_spatial_corr}. The following result characterizes the Gaussian key rate under these assumptions.

\begin{theorem}\label{thm1}
    The key rate for the memoryless ARQ protocol with {\bf Gaussian inputs} is given by:
    \begin{equation}
        \begin{split}
            C_s^{(g)}   & = \max \limits_{R_0,P\leq \bar{P}}
                            \mathbb{E} \Big\{ \big[ R_0 - log_2 \left( 1+h_eP \right) \big]^{+}\\
                        & \qquad \qquad \qquad \mathbb{I} \big(R_0 \leq \log_2 \left( 1+h_bP \right) \big) \Big\},
        \end{split}
    \label{eq_cap_general}
    \end{equation}
    \noindent for a fixed average power $P \leq \bar{P}$ and transmission rate $R_0$. $[x]^+=\max(0,x)$ and $\mathbb{I}(x) = 1$ if $x$ is true and $0$ otherwise. For the special case of spatially independent fading, i.e. $f\left(h_b,h_e\right) = f(h_b)f(h_e)$) the above expression simplifies to
    \begin{equation}
        \begin{split}
            C_s^{(i)}   & =  \max\limits_{R_0,P\leq \bar{P}}
                            \Big\{ \textrm{Pr} \big( R_0 \leq \log_2\left( 1+h_bP \right) \big)\\
                        & \qquad \qquad \qquad {\mathbb E} \big[ R_0 - \log_2 \left( 1 + h_eP \right) \big]^{+} \Big\}.
        \end{split}
    \label{eq_cap_special}
    \end{equation}
\end{theorem}

A few remarks are now in order.

\begin{enumerate}
    \item It is clear from (\ref{eq_cap_general}) that a positive secret key rate is achievable under very mild conditions on the channels experienced by Bob and Eve. More precisely, unlike the approach proposed in~\cite{Tang2007}, Theorem~\ref{thm1} establishes the achievability of a positive perfect secrecy rate by appropriately exploiting the ARQ feedback even when Eve's average SNR is higher than that of Bob.

    \item Theorem~\ref{thm1} characterizes the fundamental limit on secret key sharing and not message transmission. The difference between the two scenarios stems from the fact that the message is known to Alice {\bf before} starting the transmission of the first block, whereas Alice and Bob can defer the agreement on the key till the last successfully decoded block. This observation was exploited by our
        approach in making Eve's observations of the frames discarded by Bob, due to failure in decoding, useless.

    \item It is intuitively pleasing that the secrecy key rate in (\ref{eq_cap_special}) is the product of the probability of success at Bob and the expected value of the additional mutual information gleaned by Bob, as compared to Eve, in those successfully decoded frames.

    \item The achievability of (\ref{eq_cap_general}) hinges on a random binning argument which only establishes the existence of a coding scheme that achieves the desired rate. Our result, however, stops short of explicitly finding such optimal coding scheme and characterizing its encoding/decoding complexity. This observation motivates the development of the explicit secrecy coding schemes in Section~\ref{subsec_explicit_coding}.

    \item In the aforementioned security protocol, using a noisy feedback channel will lead to mis-synchronization between Alice and Bob. This problem can be easily overcome at the expense of a larger overhead in the feedforward channel. Alice would include all the history of received ACK/NACK in each frame. Once an ACK is received, Alice will be assured that Bob has correctly received the past history. Alice will then flush the past history and will only include the recently received ACK/NACK messages in future transmissions. Additionally, one may be tempted to assume that the noisy feedback from Bob to Eve will allow for increasing the secret key capacity. Unfortunately, Eve can easily overcome the loss of ACK bits via an exhaustive trial and error approach. More rigorously, since the ratio of feedback bits over feedforward bits is vanishingly small, the loss of ACK bits will not lead to an increase in the equivocation at Eve.

\end{enumerate}

\subsection{Spatial and Temporal Correlation}\label{subsec_spatial_corr}

One of the important insights revealed by Theorem~\ref{thm1} is the negative relation between the achievable key rate and the spatial correlation between the main and eavesdropper channels. In fact, one can easily verify that the key rate collapses to zero in the fully correlated case (i.e., $h_b=h_e$ with probability one) independent of the marginal distribution of $h_b$. In this section, we propose a solution to this problem based on a novel utilization of ``dumb antennas." The concept of dumb antennas was introduced in~\cite{Viswanath2002} as a means to create artificial channel fluctuations in slow fading environments. These fluctuations are used to harness opportunistic performance gains in multi-user cellular networks. As indicated by the name, one of the attractive features of this approach is that the receiver(s) can be oblivious to the presence of multiple transmit antennas~\cite{Viswanath2002}. We use dumb transmit antennas to \emph{de-correlate} the main and eavesdropper channels as follows. Alice is equipped with $N$ transmit antennas, whereas both Bob and Eve still have only one receive antenna. In order to simplify the presentation, we focus on the case of the symmetric fully correlated line of sight channels; whereby the magnitudes of the channel gains are all equal to one. The rest of our modeling assumptions remain as detailed in Section~\ref{subsec_system_model}. The same data stream is transmitted from the $N$ transmitted after applying an i.i.d uniform phase to each of the $N$ signals. Also, Bob is assumed to perturb its location in each ARQ frame resulting in a random and independent phase shift (from that experienced by Eve). Our multiple transmit antenna scenario, therefore, reduces to a single antenna fading wiretap channel with the following {\bf equivalent} channel gains
\begin{eqnarray*}
    g_b^{eq} = \sum\limits_{n=1}^N \left( \frac{1}{\sqrt{N}} \exp(\theta_{iR} + \theta_{iB})\right),\\
    g_e^{eq} = \sum\limits_{n=1}^N \left( \frac{1}{\sqrt{N}} \exp(\theta_{iR}+\theta_{iE})\right),
\end{eqnarray*}
\noindent where $\theta_{iB}$, $\theta_{iE}$, and $\theta_{iR}$ are i.i.d. and uniform over $[-\pi,\pi]$ that remain fixed over one ARQ frame and change randomly from one ARQ frame to the next. One can now easily see that as $N$ increases, the marginal distribution of each equivalent channel gain approaches a zero-mean complex Gaussian with unit variance (by the Central Limit Theorem (CLT)~\cite{Papoulis2001}). It is worth noting that the correlation coefficient between the two channels' equivalent power gains depends on the instantaneous channels' phases $\theta_{iB}$'s and $\theta_{iE}$'s for $i=1,\ldots,N$. It can be easily shown that, in the limit of $N\rightarrow\infty$, this correlation coefficient between the two channels power gains converges, in a mean-square sense, to zero (please refer to Appendix~\ref{app_decor_proof} for the proof). Therefore, in the asymptotic limit of a large $N$, our dumb antennas approach has successfully transformed our fully correlated line of sight channel into a symmetric and {\bf spatially independent} Rayleigh wiretap channel; whose secrecy capacity (assuming Gaussian inputs) is reported in Theorem~\ref{thm1}. The numerical results reported in the sequel (Section~\ref{subsec_numerical_results}) demonstrate that this result is not limited to line of sight channels, and that this asymptotic behavior can be observed for a relatively small number of transmit antennas.

Thus far, we have assumed that the channel gains affecting different frames are independent. This assumption renders optimal the stationary rate allocation strategy of Theorem~\ref{thm1}. In this section, we relax this assumption by introducing temporal correlation between the channel gains experienced by successive frames. Assuming high temporal correlation, if a stationary rate strategy is employed and it is less than Eve's channel capacity, all the information transmitted will be leaked to Eve. On the other hand, if the rate is much less than Bob's channel capacity, additional gains in the secrecy capacity will not be harnessed. Hence, we are going to employ a \textbf{rate adaptation} strategy in which the optimal rate used in each frame is determined based on the past history of ACK/NACK feedbacks and the rates used in previous blocks. More specifically, following in the footsteps of~\cite{Aggarwal2009}, the optimal rate allocation policy can be formulated as follows (assuming a short term average power constraint $P$ and a Gaussian input distribution).
\begin{equation}\label{eq_rate_policy}
    R_t = \arg \max \limits_{R_t}{\left\{\left(C_{s,t} + \sum\limits_{k=t+1}^{\infty}C_{s,k}\right) \Big{|} \mathbf{R}_{t-1}, \mathbf{K}_{t-1}\right\}},
\end{equation}
\noindent where
\begin{equation*}
    C_{s,t} = \textrm{Pr}\big(R_t \leq \log_2(1+h_{b, t}P)\big){\mathbb E}_{h_e} \big[ R_t - \log_2 (1+h_{e} P) \big]^{+},
\end{equation*}
\noindent where $\mathbf{R}_{t-1} = \left[R_0,\cdots,R_{t-1}\right]$ is the vector of previous transmission rates and $\mathbf{K}_{t-1} = \left[K_0,\cdots,K_{t-1}\right]$ is the vector of previously received ACKs and NACKs. The basic idea is that, after frame $(t-1)$, the posteriori distribution of $h_b$ is updated using $\mathbf{R}_{t-1}$ and $\mathbf{K}_{t-1}$. The expected secrecy rate, in future transmissions, is then maximized based on this updated distribution. It is worth noting that the above expression assumes {\bf no spatial correlation} between $h_e$ and $h_b$. This assumption represents the worst case scenario since it prevents Alice from learning the channel gains impairing Eve through the ARQ feedback. Since the channel gain is not observed directly, but through an indicator in the form of ARQ feedback, the optimal rate assignment, when the channel is Markovian, is a Partially Observable Markov Decision Process (POMDP). The solution of this POMDP is computationally intractable except for trivial cases. This motivates the following greedy rate allocation policy
\begin{equation*}
    R_t = \arg \max \limits_{R_t}{\left\{C_{s,t} \Big{|} \mathbf{R}_{t-1}, \mathbf{K}_{t-1}\right\}}.
\end{equation*}
Interestingly, the numerical results reported in Section~\ref{subsec_numerical_results} demonstrate the ability of this simple strategy to harness significant performance gains in first order Markov channels. Note that the performance of {\bf any} rate allocation policy can be upperbounded by the ergodic capacity with transmitter CSI (and short term average power constraint $P$), i.e.,
\begin{equation}\label{eq_rate_ergodic}
    C_{er} = {\mathbb E}_{h_e,h_b} \big[\log_2 (1+h_{b} P)- \log_2 (1+h_{e} P) \big]^{+},
\end{equation}
\noindent which is achieved by the optimal rate allocation policy $R_t=\log_2 (1+h_{b,t} P)$. In fact, one can view the rate assignment policy of (\ref{eq_rate_policy}) as an attempt to approach the rate of (\ref{eq_rate_ergodic}) by using the ARQ feedback to obtain a better estimate of $h_{b,t}$ after each fading block.

\subsection{Explicit Coding Schemes}\label{subsec_explicit_coding}

This section develops explicit secrecy coding schemes that allow for sharing keys using the underlying memoryless ARQ protocol with realizable encoding/decoding complexity and delay. We proceed in three steps. The first step replaces the random binning construction, used in the achievability proof of Theorem~\ref{thm1}, with an explicit coset coding scheme for the erasure-wiretap channel. This erasure-wiretap channel is created by the ACK/NACK feedback and accounts for the computational complexity available to Eve. In the second step, we limit the decoding delay by distributing the key bits over only a finite number of ARQ frames. Finally, we replace the capacity
achieving Gaussian channel code with practical coding schemes in the third step. Overall, our three-step approach allows for a useful performance-vs-complexity tradeoff.

The perfect secrecy requirement used in the information theoretic analysis does not impose any limits on Eve's decoding complexity. The idea now is to exploit the finite complexity available at Eve in simplifying the secrecy coding scheme. To illustrate the idea, let's first assume that Eve can only afford maximum likelihood (ML)
decoding. Hence, successful decoding at Eve is only possible when $R_0\leq \log_2(1+h_e P)$, for a given transmit power level $P$. Now, using the idealized error detection mechanism, Eve will be able to identify and {\bf erase} the frames decoded in error resulting in an {\bf erasure wiretap channel model}. In practice, Eve may be able to go beyond the performance of the ML decoder. For example, Eve can generate a list of candidate codewords and then use the error detection mechanism, or other means, to identify the correct one. In our setup, we quantify the computational complexity of Eve by the amount of side information $R_{\rm c}$ bits per channel use offered to it by a Genie. With this side information, the erasure probability at Eve is given by
\begin{equation}
    \epsilon=\textrm{Pr}\left(R_0-R_{\rm c}> \log_2(1+h_eP)\right),
    \label{eq_rc}
\end{equation}
\noindent since now the channel has to supply only enough mutual information to close the gap between the transmission rate $R_0$ and the side information $R_{\rm c}$. The ML performance can be obtained as a special case of (\ref{eq_rc}) by setting $R_{\rm c}=0$.

It is now clear that using this idea we have transformed our ARQ channel into an erasure-wiretap channel. In this equivalent model, we have a noiseless link between Alice and Bob, ensured by the
idealized error detection algorithm, and an erasure channel between Alice and Eve. The following result characterizes the achievable performance over this channel.

\begin{lemma}\label{erasure1}

    The secrecy capacity for the equivalent erasure-wiretap channel is
    \begin{equation*}
        \begin{split}
            C_e & =  \max\limits_{R_0, P\leq \bar{P}} \bigg\{R_0 \mathbb{E}\Big[\mathbb{I}\big((R_0 \leq \log_2(1+h_bP) \big) \\
                & \qquad \qquad \qquad (R_0 - R_c \geq \log_2(1+h_eP))\big)\Big]\bigg\}  \\
                & =  \max\limits_{R_0, P\leq \bar{P}}  \Big\{ R_0\textrm{Pr}\big(R_0 \leq \log_2 (1+h_bP), \\
                & \qquad \qquad \qquad R_0-R_{\rm c} > \log_2(1+h_eP)\big) \Big\}.
        \end{split}
    \end{equation*}
    In the case of spatially independent channels, the above expression reduces to
    \begin{equation}
        \begin{split}
            C_e & =  \max\limits_{R_0, P\leq \bar{P}} \Big\{R_0~\textrm{Pr}\big(R_0 \leq \log_2(1+h_bP)\big) \\
                & \qquad \qquad \textrm{Pr}\big(R_0-R_{\rm c} > \log_2(1+h_eP)\big)\Big\}.
        \end{split}
    \label{sp_erasure}
    \end{equation}
\end{lemma}

The proof follows from the classical result on the erasure-wiretap channel~\cite{Ozarow1984}. It is intuitively
appealing that the expression in (\ref{sp_erasure}) is simply the product of the transmission rate per channel use, the probability of successful decoding at Bob, and the probability of erasure at Eve. The main advantage of this equivalent model is that it lends itself to the explicit coset LDPC coding scheme constructed in~\cite{Bloch2006,Thangaraj2005,Thangaraj2007}. In summary, our first low complexity construction is a concatenated coding scheme where the outer code is a coset LDPC for secrecy and the inner one is a capacity achieving Gaussian code. {\bf The underlying memoryless ARQ is used to create the erasure-wiretap channel matched to this concatenated coding scheme}.

The second step is to limit the decoding delay resulting from the distribution of key bits over an asymptotically large number of ARQ blocks in the previous approach. To avoid this problem, we limit the number of ARQ frames used by the key to a finite number $k$. The implication for this choice is a non-vanishing value for the secrecy outage probability. For example, if we encode the message as the syndrome of the rate $(k-1)/k$ parity check code, Eve will be \emph{completely blind} about the key if {\em at least} one of the $k$ ARQ frames is
erased~\cite{Bloch2006,Thangaraj2005,Thangaraj2007} (Here the distilled key is the modulo-$2$ sum of the key parts received correctly). The secrecy outage probability, assuming spatially independent channels, is therefore
\begin{equation}
    P_{\rm out}= {\rm Pr}\left(\min\limits_{j\in\{1,...,k\}} \log_2
    (1+h_{e}(j)P)> R_0 - R_{\rm c}\right),
    \label{eq_prob_outage}
\end{equation}
\noindent where $h_e(1)$,...,$h_e(k)$ are i.i.d. random variables drawn
according to the marginal distribution of Eve's channel. Assuming a Rayleigh
fading distribution, we get
\begin{equation}
    P_{\rm out}=\exp\left(-\frac{k}{P}\left[2^{R_0-R_{\rm
    c}}-1\right]\right) \label{pout}.
\end{equation}
Under the same assumption, it is straightforward to
see that the average number of Bernoulli trials required to transfer
$k$ ARQ frames successfully to Bob is given by $N_0=k\exp\left(\frac{2^{R_0}-1}{P}\right)$, resulting in a key rate
\begin{equation}
R_k=\frac{R_0}{N_0}=\frac{R_0}{k}\exp\left(-\frac{2^{R_0}-1}{P}\right).
\end{equation}
Therefore, for a given $R_{\rm c}$ and $P$, one can obtain a
tradeoff between $P_{\rm out}$ and $R_k$ by varying $R_0$. Our
third, and final, step is to relax the assumption of a capacity
achieving inner code. Section~\ref{subsec_numerical_results} reports numerical results with practical coding schemes, including uncoded transmission, with a finite
frame length $n_1$. Overall, these results demonstrate the ability of the proposed protocols to achieve near-optimal key rates, under very mild assumptions, with realizable encoding/decoding complexity and bounded delay that are of practical relevance. In the next section, we introduce an ARQ-based secrecy scheme for Wi-Fi networks that builds, in principle, on these protocols.


\section{ARQ Security for Wi-Fi Networks}\label{sec_wifi}

\subsection{Wi-Fi Security: The State of the Art}\label{subsec_wifi_background}
\begin{figure*}
    \centering
        \subfigure[The encapsulation process.]{\label{fig_encap}\includegraphics[width= 0.40 \textwidth]{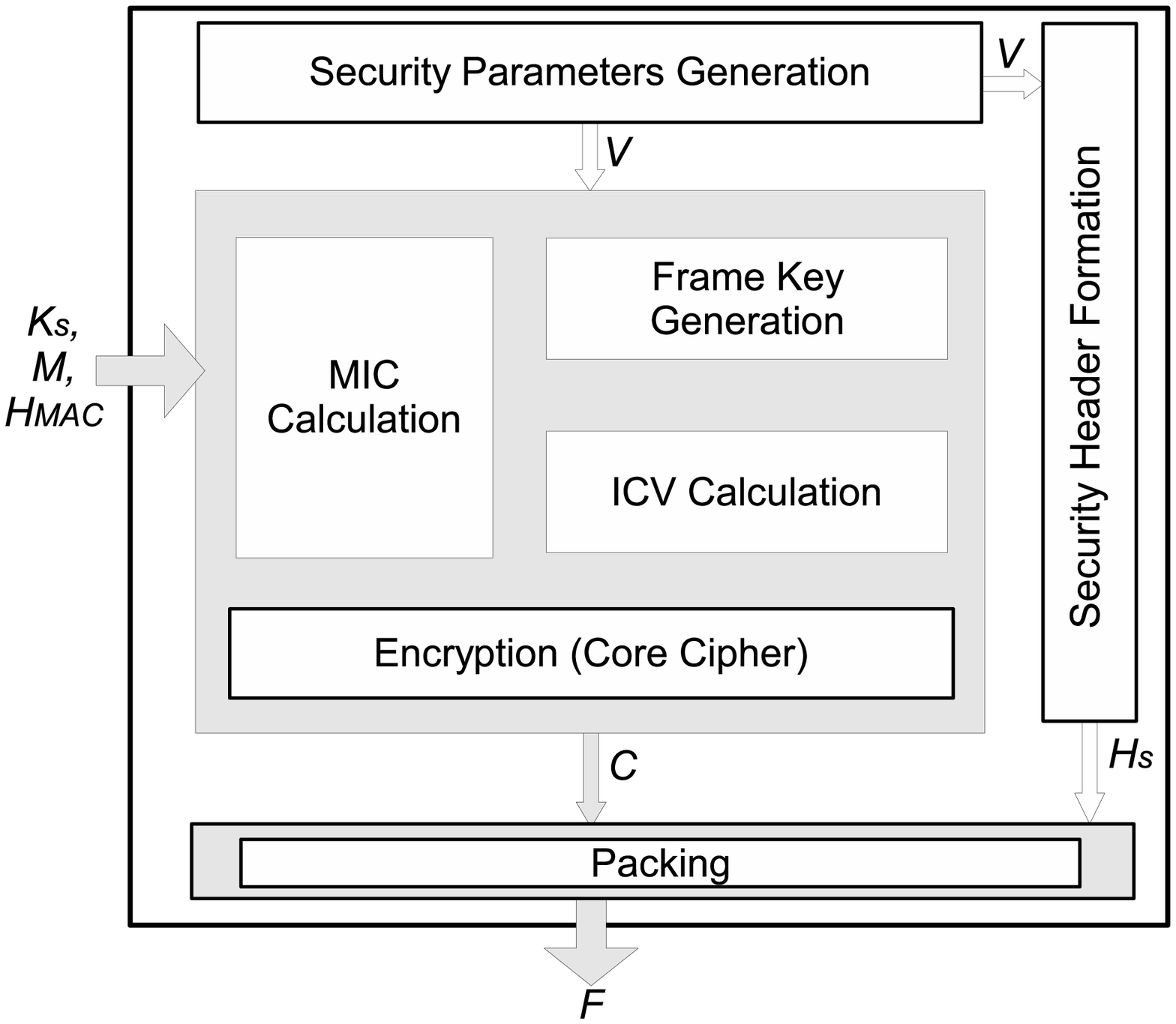}}
        \subfigure[The decapsulation process.]{\label{fig_decap}\includegraphics[width= 0.40 \textwidth]{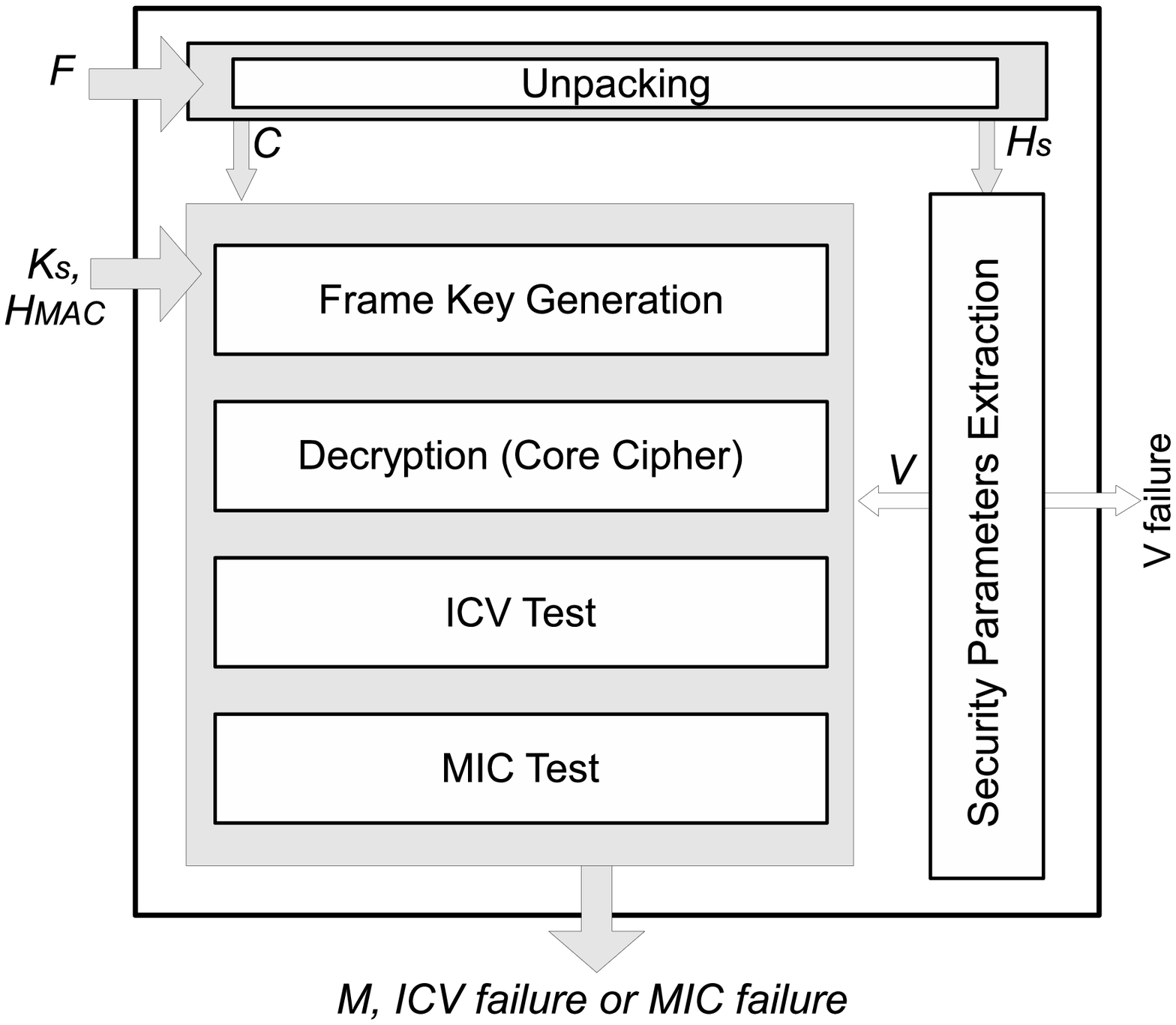}}
        \caption {WLAN-layer security functions. For a given frame, $M$ is the plaintext, $C$ is the ciphertext, and $F$ is the transmitted packet. $H_{MAC}$ and $H_s$ denote the MAC and security headers for that frame, respectively.}
    \label{fig_wlan}
\end{figure*}
Before going into the details of our design, we provide some necessary background about the existing Wi-Fi security protocols. More specifically, we describe how ``per-frame keys'' are generated and the critical dependence of all the currently-known eavesdropping attacks on weaknesses in the per-frame key generation mechanisms.

In general, the security functions of different Wi-Fi protocols could be separated into three layers, namely, an authentication layer, an access control layer and a WLAN layer~\cite{Edney2003}. In this paper, we focus only on the processes involved with encrypting and decrypting frames, that are found in the WLAN layer solely (the Wired Equivalent Privacy (WEP), the Temporal Key Integrity Protocol (TKIP), and the Counter Mode with Cipher Block Chaining Message Authentication Code Protocol (CCMP) standards). The reader is referred to~\cite{Edney2003} for details on the other two layers. We refer to the overall processes of sending and receiving frames securely as encapsulation and decapsulation, respectively. Those processes fall within WEP, TKIP (in WPA or WPA2) and CCMP (in WPA2). Figure~\ref{fig_wlan} shows two abstract schematic diagrams of frame encapsulation and decapsulation which will be useful in describing the integration of the ARQ-CROWN overlay with each of these protocols.

\subsubsection{Security at the WLAN Layer}\label{subsubsec_wifi_wlan}

The encapsulation process starts by what we refer to as ``security parameters generation", which is the first block in Figure~\ref{fig_encap}. The sole function of those generated parameters is to ensure the use of a \emph{fresh key} for each frame. In the WEP protocol, a 24-bit value, called the Initialization Vector (IV), is generated in this step. TKIP generates a similar 48-bit value, called TKIP Sequence Counter (TSC), while CCMP generates the Packet Number (PN), of length 48 bits as well.

The WEP protocol does not specify how the IV should be generated, although it recommends that the IV value should be different for each frame~\cite{Borisov2001}. In TKIP and CCMP, both the TSC or the PN are initialized by an agreed-upon value and are incremented by one for each new frame. There are two basic reasons for incrementing the TSC (or PN) versus using a random value. First, to ensure covering the entire sequence space. Second, and more importantly, to defend against replay attacks, as will be illustrated shortly. Since those parameters will be needed for decapsulation at the receiver, they are sent, \textbf{in-the-clear}, in a special security header ($H_s$) that is inserted between the frame's MAC header and the encrypted message. The remainder of the encapsulation process involves frame key generation (this is where the security parameters are combined with some secret root key, $K_s$, to obtain a key for a specific frame), encryption, adding an Integrity Check Value (ICV) and possibly a Message Integrity Check (MIC) value. We refer the reader to~\cite{Edney2003} for a comprehensive study on each of those steps.

At the receiver side (Figure~\ref{fig_decap}), the security parameters are extracted from the security header. The WEP protocol does not perform any checks on this value and directly proceeds to the next steps. However, for TKIP and CCMP, once the TSC (or the PN) is extracted from the security header, a check is performed. If the recovered TSC (PN) is less than the last received TSC (PN), the frame is considered a \emph{replayed} version of a previous frame and is \emph{discarded}. Subsequent decapsulation processes include decryption and ICV and MIC tests. Those tests serve as means to ensure that the frame has been decrypted correctly and has not been maliciously tampered with. For the purpose of this paper, we use the symbol $V$ to refer to WEP's IV, TKIP's TSC or CCMP's PN.

\subsubsection{Wi-Fi Security Attacks}\label{subsubsec_wifi_attacks}

Borisov, Goldberg, and Wagner first reported WEP design failures in~\cite{Borisov2001}. They showed that the ICV test fails to detect malicious attacks and that IV reuse allows for packet injection. Later, the first key recovery attack against WEP (the FMS attack) was presented by Fluhrer, Mantin and Shamir~\cite{Fluhrer2001} using some weaknesses of the RC4 Key Scheduling Algorithm. They also showed the recovery of the WEP key becomes much easier if some IVs that satisfy certain properties (weak IVs) were used. The KoreK chopchop attack attempted at breaking WEP using the CRC32 checksum (the ICV test)~\cite{KoreK2004}. KoreK also presented another group of attacks that do not rely on weak IVs~\cite{KoreK2004a}. A rather efficient iterative algorithm that recovers the WEP key was proposed by Klein in~\cite{Klein2008}. On the other hand, the Bittau attack made use of the fragmentation support of IEEE 802.11 to break WEP~\cite{Bittau2006}. Finally, Pyshkin, Tews, and Weinmann presented more enhancements to the Klein attack by using ranking techniques~\cite{Tews2008}. At the moment, this recent attack is considered to be the most powerful attack against WEP.

Statistical WEP attacks, e.g.~\cite{Fluhrer2001}, could, in principle, use only passive eavesdropping in order to collect a large number of frames with known IVs. However, they often use injection or replay techniques to shorten the listening time. For example, an attacker might continuously replay captured ARP (Address Resolution Protocol) request packets. Consequently, the Access Point (AP) will begin to broadcast those ARP request packets, and IVs will be generated at a higher rate. Other WEP attacks do not need a large number of IVs. Instead, they rely on injection, e.g.,~\cite{KoreK2004} or~\cite{Bittau2006}.

In 2004, weaknesses in the temporal key hash of TKIP were shown~\cite{Moen2004}. An attacker could use the knowledge of a few keystreams and TSCs to predict the Temporal Key and the MIC Key used in TKIP. Later in 2008, Tews and Beck~\cite{Tews2009} made the first practical attack against TKIP. In a chopchop-like manner, an
attacker can recover the plaintext of a short packet and falsify it within about 12-15 minutes, in a WPA network that supports IEEE802.11e QoS features. In 2009, a practical falsification attack against TKIP was proposed~\cite{Ohigashi}, in which the Beck-Tews attack was applied to a man-in-the-middle attack to target any WPA network. The latter attack takes about one minute. CCMP arguably provides robust security. However, a weakness in the nonce construction mechanism in CCMP was recently discovered~\cite{Khan2008}. A predictable PN in CCMP was shown to decrease the effective encryption key length from 128 bits to 85 bits~\cite{Khan2008}.

In summary, the previously mentioned attacks rely on collecting a large number of ciphertext along with the corresponding security parameters {\em which are sent in-the-clear}, whether through passive eavesdropping or innovative active techniques. As detailed in the following section, the ARQ-CROWN overlay solves this problem by exploiting the opportunistic secrecy principle resulting from the wireless multipath fading environments.

\subsection{ARQ-CROWN: An Overview}\label{subsec_crown_overview}


ARQ-CROWN is designed for Wi-Fi networks operating in infrastructure mode that may use any of the IEEE802.11 security protocols, i.e., WEP, TKIP or CCMP for encryption. The network is composed of one AP and $L$ clients, in the presence of one attacker. The AP and all clients follow the ARQ mechanism adopted in the IEEE 802.11
standard, i.e., for each transmitted frame, the receiver acknowledges the receipt of that frame through a short ACK message. We assume disabled retransmissions, i.e., if a timeout event occurs at the transmitter (the data frame or the ACK message were lost), it simply discards the
current frame and moves to further transmissions\footnote{The analysis provided in this paper could be easily extended to the case of enabled retransmissions.}.

Key management and re-keying policies are aspects that fall outside the scope of this paper. For this reason, we assume that once a wireless client is authenticated and has gained access to the network, it shares root keys with the AP. From the simplest setting of one-key-for-all in the WEP protocol, to a rather complicated key hierarchy in WPA and WPA2, our discussion would be on a per-frame basis. Hence, we assume that, for each frame, the client and the AP agree on which key is used to encapsulate/decapsulate this frame. Throughout the sequel, this secret key is referred to as $K_{s}$.
\begin{figure*}
    \centering
        \subfigure[The encapsulation process.]{\label{fig_arq_encap}\includegraphics[width= 0.40 \textwidth]{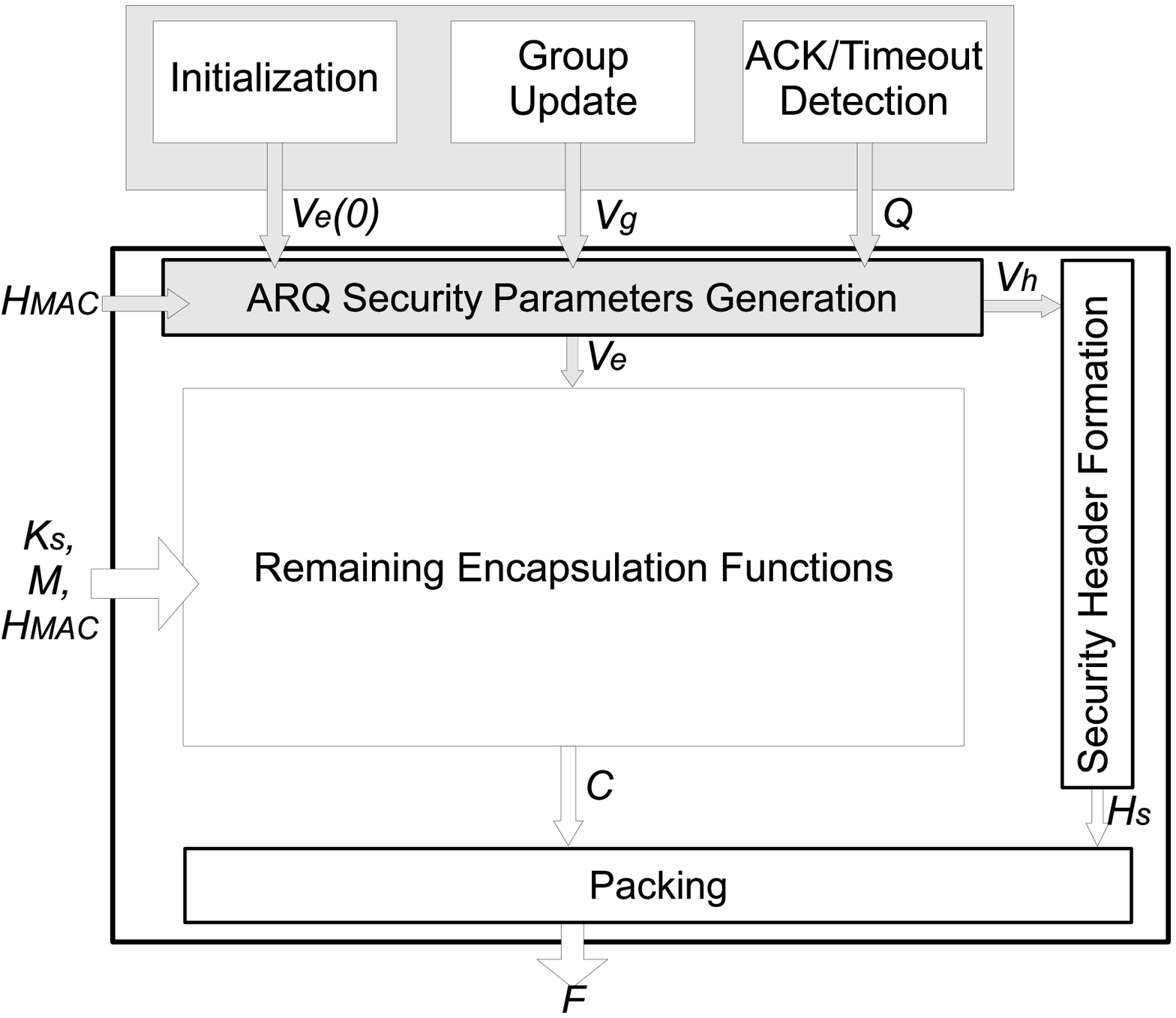}}
        \subfigure[The decapsulation process.]{\label{fig_arq_decap}\includegraphics[width= 0.40 \textwidth]{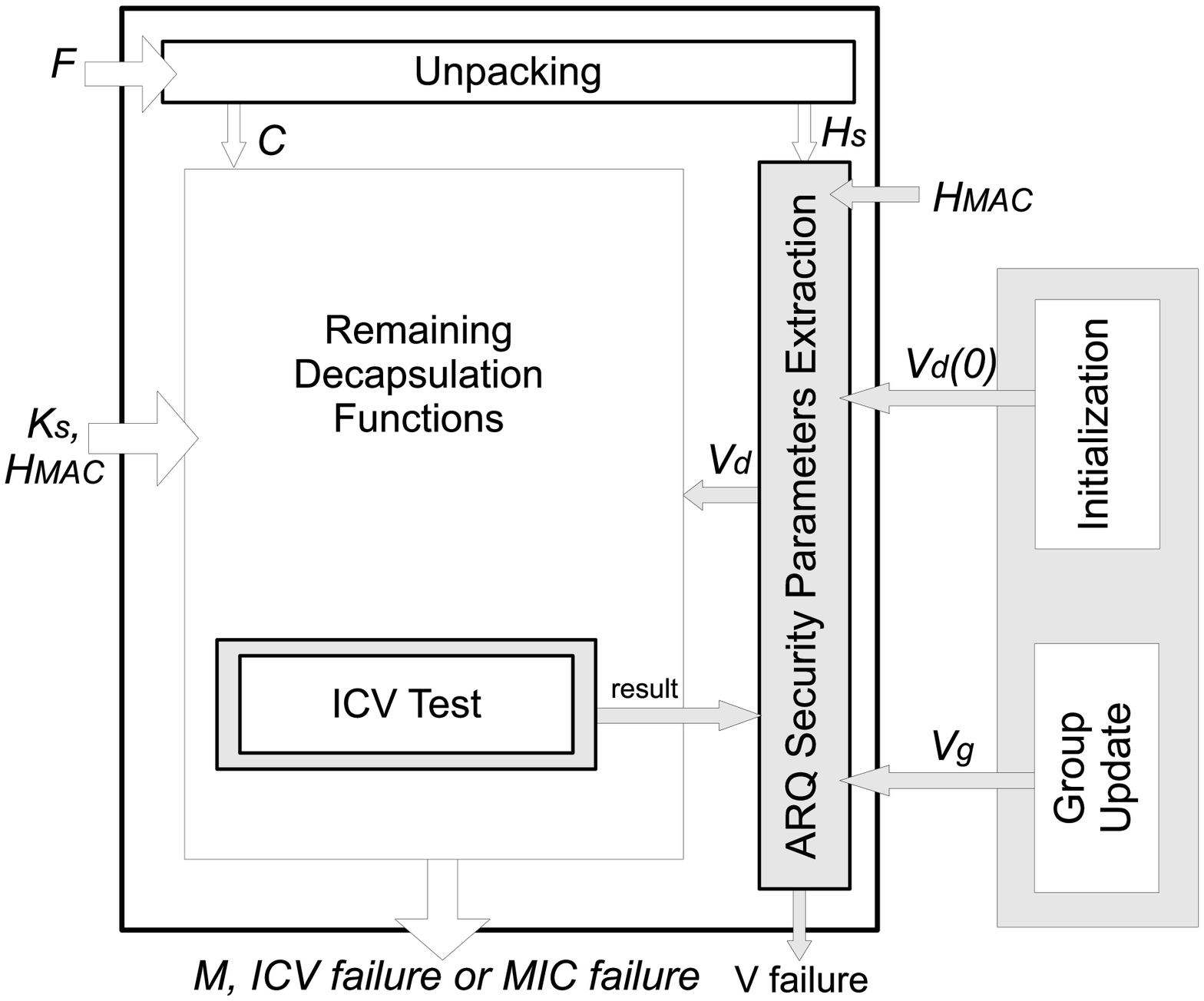}}
        \caption {WLAN-layer security functions incorporating the ARQ-CROWN overlay. The shaded blocks represent ARQ-CROWN modifications}
    \label{fig_arq_wlan}
\end{figure*}
In the proposed ARQ-CROWN overlay, we transform the $V$ values of different frames into additional private
keys that are shared among the legitimate nodes. ARQ-CROWN entirely focuses on the $V$ value of each frame, leaving the secret root key, $K_s$, unaltered. Figure~\ref{fig_arq_wlan} shows the modified WLAN layer when overlaid by ARQ-CROWN. The figure shows three new separate modules that run independently from the encapsulation and decapsulation processes; namely, an initialization module, an ACK/Timeout detection module and a group update module. Those modules interact solely with the security parameters generation and extraction blocks that are modified to incorporate ARQ security. Outputs of those steps are fed to the remaining functional blocks of encapsulation and decapsulation, which remain exactly the same as in the original standards. For ease of presentation, we begin by using a simple three-node network model. In this network, Alice corresponds to one legitimate client, Bob corresponds to the AP and Eve is a malicious attacker. We later show how to extend our scheme to secure multicast flows.

The initialization module works on letting Alice and Bob agree on an initial value, $V_0$, that will be later used in securing unicast flows in the Alice-Bob and Bob-Alice directions. It runs, only once, after Alice is associated and authenticated and before data ports are open. In essence, the process is similar to the one described in Section~\ref{subsec_system_model} but with some modifications that better utilize the MAC layer of the IEEE 802.11 standard and that take into account dealing with an active eavesdropper, as will be clear with further discussion. Once this initialization phase is complete, secure data communication is allowed. The ACK/Timeout detection module runs during open data sessions. It works on deciding on the status of each transmitted unicast frame, which is referred to as $Q$. This status helps both Alice and Bob update the $V$ values for the unicast frames they exchange, for \textbf{each} transmitted frame. Finally, the group update module allows for securing multicast data. In the following section, we show how each of those modules operate and rigorously analyze their security.

\subsection{ARQ-CROWN: Operation and Security Analysis}\label{subsec_crown_details}

\subsubsection{The Initialization Phase}\label{subsubsec_crown_init}
\begin{figure}
\centering \epsfig{file=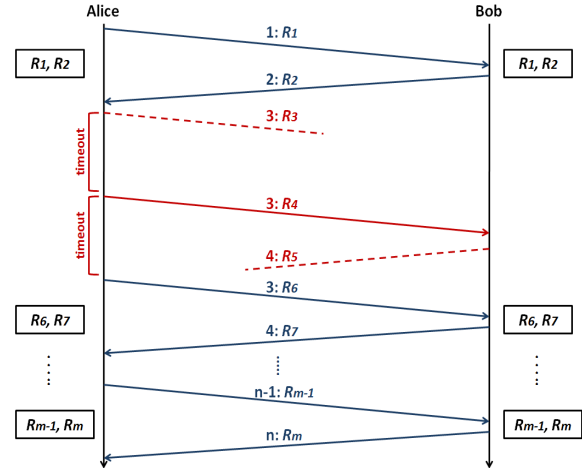,
width=0.45\textwidth} \caption{The ARQ-CROWN initialization phase.}\label{fig_linksetup}
\end{figure}
The initialization phase works as illustrated in Figure~\ref{fig_linksetup}. First, Alice transmits an initialization frame, carrying a sequence number $1$ and random number $R_1$, and starts a timer. Once Bob
receives this frame, he replies with another initialization frame, carrying a sequence number $2$, and another random number $R_2$. If Alice receives this frame before a timeout event occurs, she stores the pair $(R_1,R_2)$ for later use, and transmits another initialization frame with sequence number $3$ and a new random
number $R_3$. Otherwise (a timeout event occurs), Alice discards $R_1$, and transmits another initialization frame with sequence number $1$ and a new random number $R_3$. The process continues till Alice has stored $n$ initialization random values. On the other side, Bob keeps on responding to each initialization frame he gets
with a sequence number incremented by one, and a newly generated random number. However, he stores only the last pair it has for any given sequence number. The length of each transmitted random number is $24$ bits if WEP is used, or $48$ bits otherwise. Finally, the initial value, $V_0$, is the modulo-2 sum of the random number pairs successfully received by \textbf{both} Alice and Bob.

The security of this protocol in the presence of a passive Eve directly builds on the results provided in Section~\ref{subsec_explicit_coding}. More specifically, as Eve becomes completely blind about $V_0$ if she misses one of the values constituting $V_0$, the probability of secrecy outage in our case (corresponding to (\ref{eq_prob_outage})) is
\begin{equation}\label{eq_p0}
    P_{0} = \prod_{i\in \mathcal{A}} (1-\gamma_{AE_i}) \prod_{j\in \mathcal{B}} (1-\gamma_{BE_{j}}),
\end{equation}
\noindent where $\mathcal{A}$ and $\mathcal{B}$ are the sets of time indices that correspond to the frames stored by Alice and Bob, respectively. $\gamma_{AE_1}, \ldots, \gamma_{AE_{n-1}}$ denote the frame loss probabilities in the Alice-Eve channel whereas $\gamma_{BE_2}, \ldots, \gamma_{BE_{n}}$ denote the frame loss probabilities in the Bob-Eve channel. All of those probabilities are random variables that are independently and identically distributed according to Eve's channels' distributions. Since the size of each of $\mathcal{A}$ and $\mathcal{B}$ is $n/2$. It is evident that, as $n$ increases, $P_0$ decreases and we achieve better security gains, at the expense of a larger delay in the initialization phase.

On the other hand, if Eve is active, she will be capable of injecting or replaying initialization frames, since they are \emph{not encrypted}. However, any injection or replay attempt will cause a disagreement between Alice and Bob on $V_0$. We will later show that if Alice and Bob do not agree on $V_0$, they will not be able to exchange any data frames. Consequently, a replay or injection attack directly corresponds to a Denial of Service (DoS) attack. We finally note that in the case of using the WEP protocol, the initialization frames, being un-encrypted, reveal no information about the secret key, $K_s$, and thus cannot be used in any statistical WEP attack.

\subsubsection{Securing Unicast Data}\label{subsubsec_crown_unicast}

Right after initialization, our protocol works on updating the $V$ values, used to encapsulate each unicast data frame sent on the Alice-Bob and Bob-Alice channels. To illustrate, first consider the $i^{th}$ data frame to be securely transmitted, using any security protocol, from Alice to Bob. Alice
starts by generating a random number (of length $24$ if WEP is used, or $48$ bits otherwise) referred to as the
header-V, $V_h(i)$. The ARQ-CROWN protocol must not use two consecutive equal header-V's. This property will be shown to be useful for defending against replay attacks. This value, $V_h(i)$, is put in the frame's security header, according to the specifications of the security protocol used. However, \textbf{unlike} the standards, the value used by ARQ-CROWN in encapsulating the frame, denoted by $V_e(i)$, is the \textbf{modulo-2 sum} of
the current header-V, $V_h(i)$, and all of the header-V's previously transmitted by Alice and successfully received by Bob. The update equation for $V_e$ is then
\begin{equation}\label{eq_vesimple}
    V_e(i) =
        \begin{cases}
            V_h(i) \bigoplus V_e(i-1),&\text{if}~Q(i-1) = 1,\\
            V_h(i) \bigoplus V_e(i-1) \bigoplus V_h(i-1),&\text{otherwise,}
        \end{cases}
\end{equation}
\noindent where $Q(i)=1$ if Alice received an ACK for the $i^{th}$ transmitted frame, $Q(i)=0$ otherwise. This status is obtained through the ACK/Timeout detection module running at Alice (Figure~\ref{fig_arq_encap}). The initial value for this algorithm is set by the agreed-upon $V_0$ of the initialization phase, i.e., $V_e(0) = V_0$, while $V_h(0) = 0$. Similarly, when Bob receives the $i^{th}$ frame, he first extracts $V_h(i)$ from the security header, and then performs a check. If $V_h(i) = V_h(i-1)$, Bob discards the frame and treats it as a sign of a replay attack. If not, Bob attempts to decapsulate the frame with $V_d(i)$,
\begin{equation}\label{eq_vd1}
    V_d(i) = V_h(i) \bigoplus V_d(i-1),
\end{equation}
where $V_d(0) = V_0$. If decryption fails (an ICV failure occurs), this would be due to an erasure of the $(i-1)^{th}$ ACK. Bob then goes through another decryption attempt, after excluding $V_h(i-1)$ from the sum, i.e., with $V_d(i) = V_h(i) \bigoplus V_d(i-1) \bigoplus V_h(i-1)$. Another failure in decryption is treated as a sign of an attack and countermeasures could be invoked (the reason behind this will become clear in the security analysis to follow). Following this protocol, Alice and Bob perfectly agree on the $V$ values used for each frame. We avoid any mis-synchronization that could happen due to the loss of an ACK frame; without any additional feedback bits (as opposed to Section~\ref{subsec_main_result}). The unicast flow from Bob to Alice could be secured in the same manner illustrated above.

We now analyze the security of this phase. In our scheme, the collected traffic by a passive Eve becomes useful for any attack depending on Eve's ability to correctly compute $V_{e}$ for each captured frame. To achieve this, Eve first has to correctly compute $V_0$, in the initialization phase between Alice and Bob. This happens with probability $P_0$ (as given in (\ref{eq_p0})). Afterwards, for each captured frame, Eve has to keep track of \textbf{all} the previously acknowledged data frames preceding that frame. Eve becomes, again, completely blind if she misses a single acknowledged frame. Based on this observation, we let $u$ denote the total number of data frames that Eve can correctly compute their $V_e$, i.e., the \emph{useful} frames for Eve. If $\gamma_{AE} = \gamma_{AB} = \gamma_{E}$ for all time indices, the expected number of such frames is upper-bounded by
\begin{equation}\label{eq_eu}
    \mathbb{E}[u]\leq\frac{\mathbb{E}[\gamma\acute{}_{E}]^{n+1}-\mathbb{E}[\gamma\acute{}_{E}]^{N+1}}
    {\mathbb{E}[\gamma_{E}]},
\end{equation}
where $\gamma\acute{}_{E} = 1 - \gamma_{E}$, $n$ is the total number of initialization frames constituting $V_0$ and $N$ is the unicast data session size. As shown in Eq. (\ref{eq_eu}), a slight increase of the number of initialization frames results in a significant decrease in the number of useful frames for Eve in each session. This has a direct impact on the feasibility of many attacks, especially the statistical WEP attacks, e..g.~\cite{Fluhrer2001}, as those depend on collecting a large number of IVs ($V_e$'s in the
ARQ-CROWN case) to run efficiently.

We now consider the case of an active Eve. For the
unicast flow from Alice to Bob, Eve could use Alice's MAC
address to inject or replay data frames of her choice, or use Bob's MAC address to inject ACK messages to confuse Alice. However, any injected or replayed frame will lead to
mis-synchronization between Alice and Bob. This will be detected by
Bob through two successive ICV failures. As we already mentioned,
Bob would treat this as a sign of an attack and countermeasures
could follow. The most straightforward countermeasure is to change
the keys of the whole network or of the attacked sessions. Still,
the history of $V$ values built up thus far could be used after
invoking countermeasures through fast means of ``re-synchronization" as will be later discussed.

Frame interception (jamming), in general, is often used as part of phishing and MITM attacks. Additionally, when ARQ-CROWN is deployed, interception could be used to delay the key update process for a certain data flow in the network. Defending against those attacks requires additional modifications, which are outlined in Section~\ref{subsec_crown_discussion}.

\subsubsection{Securing Multicast Traffic}\label{subsubsec_crown_multicast}

Thus far, our discussion was limited to unicast sessions. Since
multicast frames are not ACKed, the previously demonstrated scheme cannot be used to secure these frames. Our scheme for multicast traffic goes as follows: Whenever a client subscribes to a
multicast group, $g$, the AP sends a new random value, $V_{g}$, to
every associated client that belongs to this group along with an ID for this $V_g$ value (the updates can be periodic or triggered based on group membership changes). Those values are
transmitted to each client over its secure pairwise link with the
AP, i.e., as \emph{encrypted} frames. Once the AP makes sure that
all clients in the group have received $V_{g}$, through individual
ACKs, the AP uses this value to compute $V_{e_{g}}$, that will be used for encapsulating each upcoming multicast frame, within this group, i.e.,
\begin{equation}\label{eq_arq_wep_01}
    V_{e_{g}}(i) = V_{h}(i) \bigoplus V_{g}.
\end{equation}
\noindent where $V_{h}(i)$ is a random header-V as illustrated before. $V_h(i)$ and the ID of the used $V_g$ are sent in the security header of the multicast frame. Similarly, for members of a particular multicast group $g$, a client uses the recovered information from the security header to compute $V_{d_{g}}(i)$ and decapsulate any multicast frame addressed to this group. Any failure in decryption (ICV test failure) is treated as a sign of
attack.

Finally, in order to defend against replay attacks, the AP should not use repeated $V_h$
values within the lifetime of a certain $V_g$. Similarly, whenever a
client receives a multicast frame, it must check for this condition and treat repeated $V_h$'s as a sign of attack.

Using this ARQ-CROWN multicast overlay, a passive Eve cannot make use of any of
the multicast frames, as secure pairwise links are used to
incorporate hidden and periodically-updated values into multicast
$V_{e}$'s. On the other hand, an active Eve is not capable of
injecting or replaying any of the multicast frames, as any replay or
injection attempt would lead to a decryption failure at the
legitimate recipients. Finally, for WPA and WPA2, since there is a different group key for each
multicast group and that is updated with group membership changes, our
proposed multicast approach fits nicely within their framework and
increases their security. For the WEP case, which uses a shared key
for all multicast groups, our group-V updates add a natural way for group membership handling.
This gives an additional security advantage for the WEP case, without having to change the secret root key, $K_{s}$.

\subsection{Discussion}\label{subsec_crown_discussion}

The enhanced security, offered by our scheme, is mostly evident in the case of WEP. In particular, using the ARQ-CROWN overlay, any statistical WEP attack would require a substantially longer listening time before launching the attack; which makes such attacks virtually impossible. This is demonstrated by the experimental results of Section~\ref{subsec_experimental}. It is worth nothing that in order for Eve to have a potential use of any unicast session, she has to be present from the \textit{beginning} of this session. Also, our analytical estimate of the lower bound on the number of useful frames for Eve (Eq. (\ref{eq_p0})) implicitly assumes that Eve is totally capable of tracking ACKs, i.e., she \textit{perfectly} knows the status of each unicast frames. In practice, especially in large networks where channel conditions could be relatively worse, such knowledge is not perfect which causes more confusion at Eve's side.

One can envision several enhancements for the basic implementation presented here. First, setting the timeout periods in the ARQ-CROWN initialization phase should be carefully designed so as to defend against MITM attacks and at the same time keep the initialization delay within a practically acceptable range. A related point is to analyze the ACK/timeout events at the legitimate senders to detect anomalies in the behavior of the connected nodes for better detection of frame interception (jamming). Second, in order to reduce the overhead of the initialization phase, the legitimate nodes can use the current history for future sessions. Upon disassociation, the AP and any legitimate client can store the last point in their ARQ-history, and build up on it in newer sessions instead of going through new initialization phases. This way, the additional link setup delay imposed by the ARQ-CROWN overlay is minimized and security is enhanced at the expense of additional negligible memory at both sides. This is especially useful for designing seamless handoff mechanisms for Wi-Fi networks as this information can be transferred to the new AP using the IEEE 802.11f protocol. Finally, through small modifications, the ARQ-CROWN overlay could be further extended to secure the secret root keys to provide more security. The ARQ-CROWN overlay could also be used for security at layers higher than the MAC layer, using the same underlying principles.


\section{Numerical and Experimental Results}\label{sec_results}

\subsection{Numerical Results}\label{subsec_numerical_results}
\begin{figure}
\centering
\includegraphics[width=0.5\textwidth]{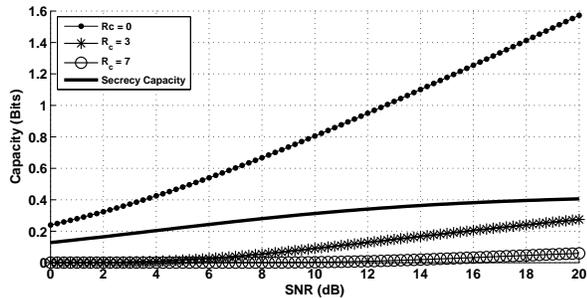}
\caption{$C_s$ and $C_e$ against SNR for $R_c =\left(0,3,7\right)$. \label{fig1}}
\end{figure}
Throughout this part, we focus on the symmetric scenario where
${\mathbb E}\left(h_b\right)={\mathbb E}\left(h_e\right)$ = 1. We further assume Rayleigh fading channels, for both Bob and Eve. Assuming spatially and temporally independent channels, the achievable secrecy rate in
(\ref{eq_cap_special}) becomes
\begin{equation}
    \begin{split}
        C_s & = \max\limits_{R_0} \exp\left(-\frac{2^{R_0}-1}{P}\right) \\
            & \qquad \qquad \left\{R_0-\frac{\exp\left(1/P\right)}{\log_e\left(2\right)}\left[E_{\rm i}
                \left(1/P\right)-E_{\rm i}\left(2^{R_0}/P\right)\right]\right\},
    \end{split}
\end{equation}
\noindent where $E_{\rm i}\left(x\right)=\int_{x}^{\infty}\exp\left(-t\right)/t\,dt$. Figure~\ref{fig1} gives the variation of $C_s$ and $C_e$ (as given in (\ref{sp_erasure})) with SNR under different constraints on the decoding capabilities of Eve, captured by the genie-given side information, $R_c$. It is clear from the figure that $C_e$ can be greater than $C_s$ for certain $R_c$ and SNR values. For instance, in the case of $R_c=0$, a
packet received in error at Eve will be discarded {\bf without any further attempts at decoding}. Therefore, the secrecy rate becomes $R_0$, which is larger than that used in (\ref{eq_cap_special}); $C_s(i)= R_0 -
\log_2(1+h_e(i)P),$ where $C_s(i), h_e(i)$ are the instantaneous secrecy rate, and Eve's channel power gain, respectively. Averaging over all fading realizations, we get a greater $C_e$ than $C_s$. It is worth noting that, under the assumptions of the symmetric scenario and the Rayleigh fading model, the scheme proposed in~\cite{Tang2007} is not able to achieve any positive secrecy rate (i.e., probability of secrecy outage is one).
\begin{figure}
\centering
\includegraphics[width=0.5\textwidth]{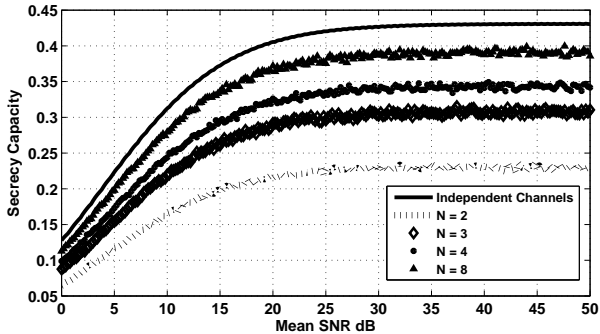}
\caption{The key rates using $N$ dumb antennas, assuming fully correlated exponential channel gains. \label{fig5_dumb}}
\end{figure}
The role of dumb antennas in increasing the secrecy capacity of spatially correlated ARQ channels is investigated next. In our simulations, we assume that the channel gains are fully correlated, but the channel phases are independent. The independence assumption for the phases is justified as a small change in distance between Bob and Eve in the order of several electromagnetic wavelengths translates to a significant change in phase. Under these assumptions, it is easy to see that with one transmit antenna the secrecy capacity is zero. In Figure~\ref{fig5_dumb}, it is shown that as the number of antennas $N$ increases, the secret key rate approaches the upper bound given by~(\ref{eq_cap_special}), which assumes that the main and eavesdropper channels are independent. The same trend is observed assuming chi-square distribution with different degrees of freedom (the figures were omitted to avoid redundancy).
\begin{figure}
\centering
\includegraphics[width=0.5\textwidth]{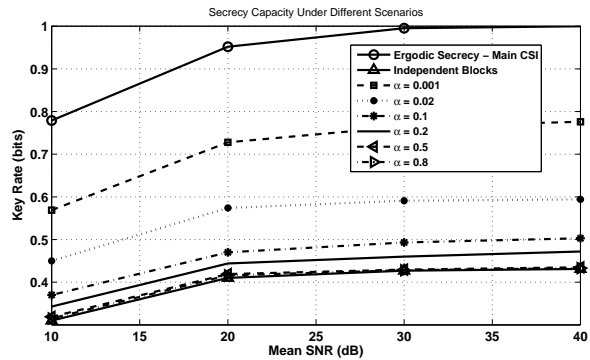}
\caption{The achievable key rates using the greedy scheme under different temporal correlation coefficient $\alpha$. \label{fig9_temp_corr}}
\end{figure}
Figure~\ref{fig9_temp_corr} reports the performance of the greedy rate adaptation algorithm for temporally correlated channels. The channel is assumed to follow a first order Markov model:
\begin{equation*}
    g(t) = (1 - \alpha) g(t-1) + \sqrt{2\alpha - \alpha^2} w(t)
\end{equation*}
\noindent where $w(t)$ is the innovation process following $\mathcal{CN}(0,1)$ distribution. As expected, it is shown that as $\alpha$ decreases, the key rate increases. For the extreme points when $\alpha = 0$ or $\alpha = 1$, we get an \textbf{upper bound}, which is the ergodic secrecy under the main-channel transmit CSI assumption, and a \textbf{lower bound}, which is the ARQ secrecy capacity in case of independent block fading channel, respectively.

\begin{figure}
\centering
\includegraphics[width=0.5\textwidth]{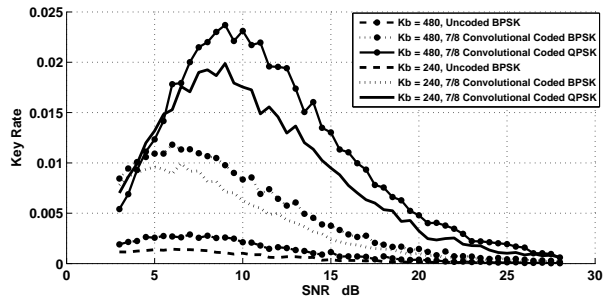}
\caption{The key rates required to obtain an outage of $10^{-10}$
against SNR for different packet sizes, $K_b=240$ and $480$ bits,
and different modulation schemes. \label{fig4}}
\end{figure}

Finally, we turn our attention to the delay-limited coding constructions proposed in Section~\ref{subsec_explicit_coding}. In Figure~\ref{fig4}, we relax the optimal channel coding assumption and plot key rates for practical coding schemes and finite frame lengthes (i.e., finite $n_1$). The code used in the simulation is a punctured convolutional code derived from a basic $1/2$ code with a constraint length of $7$ and generator polynomials $133$ and $171$ (in octal). We assume that Eve is genie-aided and can correct an additional $50$ erroneous symbols (beyond the error correction capability of the channel code). Note that the transmission rate is fixed and is independent of the SNR. Therefore, a low SNR means more transmissions to Bob and a consequent low key rate. As the SNR increases, while keeping the transmission rate fixed, the key rate increases. However, increasing the SNR also means an increased ability of Eve to correctly decode the codeword-carrying packets. This explains why the key rate curves a peak and then decays with SNR. We also observe that, for a certain modulation and channel coding scheme, reducing the packet size increases the probability of correct decoding by Bob and, thus, decreases the number of transmissions. However, it also increases the probability of correct decoding by Eve and the overall effect is a decreased key rate.

\subsection{Experimental Results}\label{subsec_experimental}

Our experiments are conducted with a modified version of the Madwifi driver that has ARQ-CROWN capabilities. All of our testbed nodes are Dell Latitude D830 laptops that are equipped with Atheros-based D-Link DWL-G650 WLAN cards. All traffic is generated using Netperf~\cite{netperf}.

\subsubsection{Security}


\begin{figure}
    \centering
        \subfigure[$\mathbb{E}{[}\gamma_{AB}{]} = 0.005,\mathbb{E}{[}\gamma_{BA}{]} = 0.009$ and $\mathbb{E}{[}\gamma_{AE}{]} = 0.004$.]{\label{secrecy_results1}\includegraphics[width= 0.35 \textwidth]{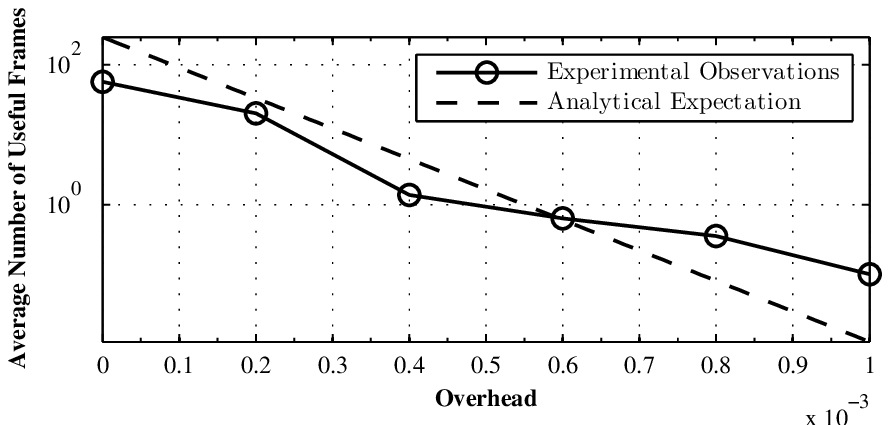}}
        \subfigure[$\mathbb{E}{[}\gamma_{AB}{]} = 0.01,\mathbb{E}{[}\gamma_{BA}{]} = 0.01$ and $\mathbb{E}{[}\gamma_{AE}{]} = 0.02$.]{\label{secrecy_results2}\includegraphics[width= 0.35 \textwidth]{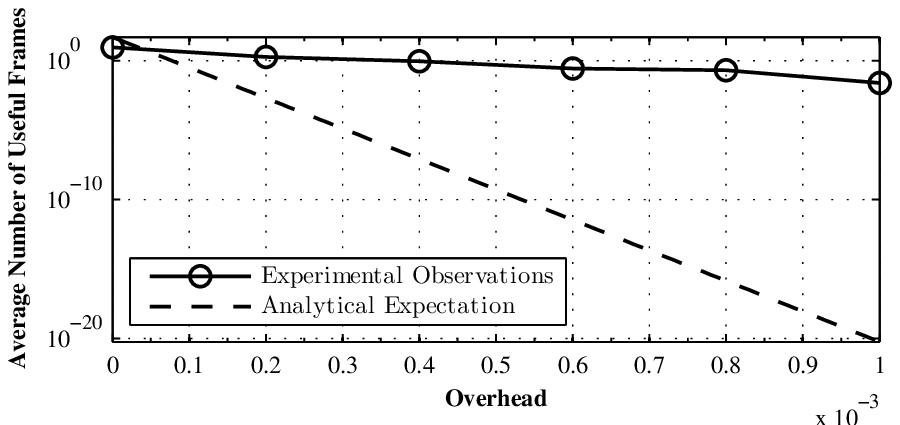}}
        \caption {The average number of useful frames at Eve.}
    \label{fig_secrecy_results}
\end{figure}

One-way traffic was generated between a client node (Alice) and the AP (Bob) in
the presence of one eavesdropper (Eve). Eve's driver was equipped with
the ARQ-CROWN algorithms, i.e. Eve calculates $V_e$ for each frame based on the captured traffic.
Two experiments were launched in different environments. In the
first experiment, Eve had relatively better channel conditions, as compared to Bob,
while in the second, the situation was reversed. We compared the $V_e$ values that Eve and Bob obtained for each frame, and calculated the number of useful frames for Eve (with different numbers of initialization frames).

The results are reported in log scale in Figure~\ref{fig_secrecy_results}. For both experiments, the data session size is taken to be 100000 frames. The large disagreement between the analytical estimates (evaluated as given in~\ref{eq_eu}) and the experimental results in Figure~\ref{secrecy_results2} is due to the very small average number (up to $10^{-20}$) of useful frames when the channel conditions are against Eve, which requires an \textbf{infeasible} experiment duration to be captured in practice. These results can be used to estimate the required time for Eve to capture a total of $1.5$ million useful frames that is typically required to launch a combined form of the FMS and KoreK attacks (~\cite{aircrack}). Under the original WEP operation, we assume that Eve needs \textbf{$10$ minutes} to gather such traffic using passive eavesdropping only. Based on this estimate, using ARQ-WEP protocol extends the required average listening time for Eve to \textbf{$1.24$ years} and \textbf{$5.07$ years}, for the first and second experiments, respectively, using only an initialization overhead of $0.001$. Note that under ARQ-CROWN operation, Eve cannot use any active techniques to reduce the listening time. For TKIP and CCMP, the decreased number of useful frames at Eve hampers her ability to exploit the weaknesses that were discussed in Section~\ref{subsubsec_wifi_attacks}.

\subsubsection{Throughput}

\begin{figure}
    \centering
        \subfigure[WEP is used for encryption.]{\label{fig_thruput_01}\includegraphics[width= 0.32 \textwidth]{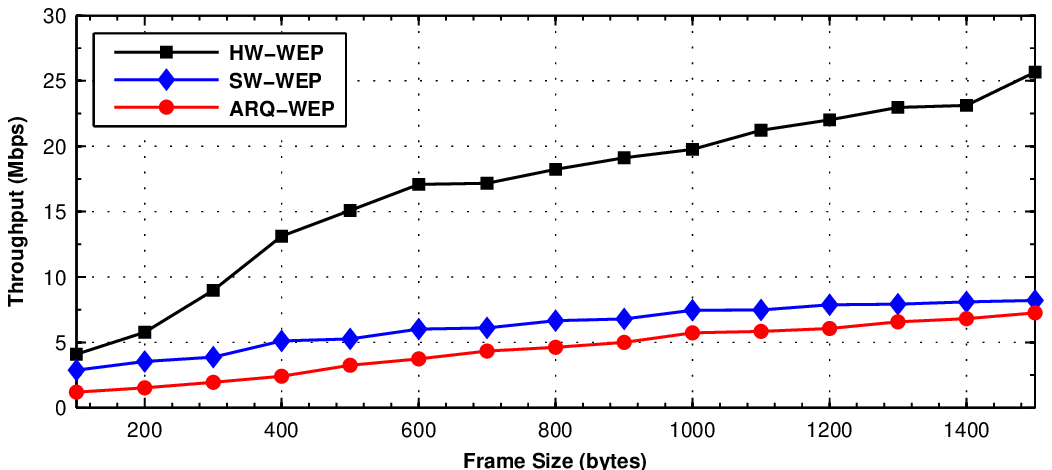}}
        \subfigure[TKIP is used for encryption.]{\label{fig_thruput_02}\includegraphics[width= 0.32 \textwidth]{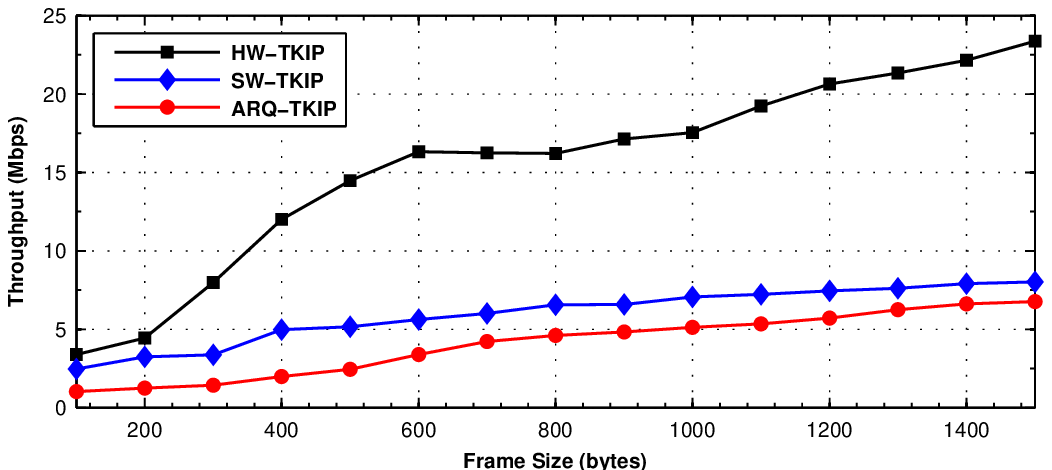}}
        \subfigure[CCMP is used for encryption.]{\label{fig_thruput_03}\includegraphics[width= 0.32 \textwidth]{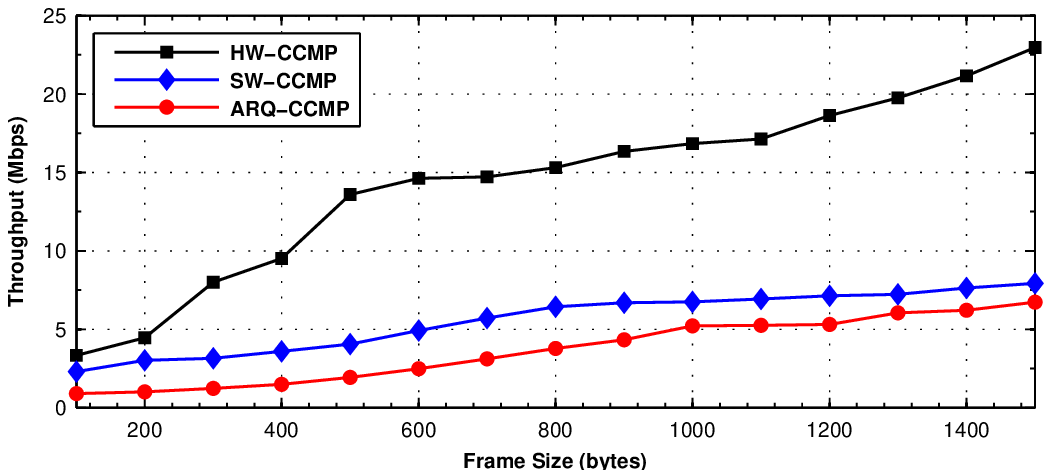}}
        \caption {Network throughput for TCP flows with different security protocols.}
    \label{fig_thruput_results}
\end{figure}

Here we compare the performance of the proposed ARQ-CROWN overlay with the baseline software implementations of WEP, TKIP, and CCMP in the Madwifi driver. To obtain a measure of performance if the proposed ARQ-CROWN overlay was implemented in hardware, we also include the results of all hardware implementations. Figure~\ref{fig_thruput_results} reports the aggregate network throughput for TCP flows, with different packet sizes, for WEP, TKIP, and CCMP. One can see that using the ARQ-CROWN on top of WEP (ARQ-WEP) results in a throughput degradation of 11.57\% over the Madwifi software implementation of WEP (SW-WEP), for a packet size of 1500 bytes. The corresponding degradation for TKIP and CCMP is 15.61\% and 15.26\%, respectively. This quantifies the processing overhead of ARQ-CROWN operation (as described in Section \ref{subsubsec_crown_unicast}). As the packet size increases, the overhead introduced by the ARQ-CROWN decreases, as it is amortized over a larger packet size.

\section{Conclusions}\label{sec_conclusions}

This paper developed a unified framework for sharing secret keys using existing ARQ protocols. The underlying idea is to distribute the key bits over multiple ARQ frames and then use the authenticated ACK/NACK feedback to create an equivalent degraded channel at the eavesdropper. Our information theoretic foundations established the achievability of non-zero secrecy rates even when the eavesdropper is experiencing a higher average SNR than the legitimate receiver and shed light on the structure of optimal ARQ secrecy protocols. It is worth noting that our approach does not assume any prior knowledge about the instantaneous CSI; only prior knowledge of the average SNRs seen by the eavesdropper and the legitimate receiver are needed. Our secrecy capacity characterization revealed the negative impact of spatial correlation and the positive impact of temporal correlation on the achievable key rates. The former phenomenon was mitigated via a novel ``dumb antennas" technique, whereas the latter was exploited via a greedy rate adaptation policy. Furthermore, low complexity secrecy coding schemes were constructed by transforming our channel to an erasure wiretap channel which lends itself to explicit coset coding approaches. Building on this solid foundation, we developed a novel approach for ARQ security in Wi-Fi networks (i.e., ARQ-CROWN). Our ARQ-CROWN overlay is shown to offer provable information theoretic confidentiality guarantees which complement the security measures provided by the underlying WEP, WPA, and WPA2 protocols. These claims were validated by experimental results, obtained from our prototype, which illustrate the ability of ARQ-CROWN to mitigate all known eavesdropping attacks, whether active or passive, at the expense of a throughput loss in the order of $10\%$--$15\%$ using software encryption.

{\bf The most interesting part of our work is, perhaps, the demonstration of the utility of information theoretic security concepts in securing state of the art wireless networks.} In our opinion, the success of such concepts in practice will depend critically on the ability to apply them to complement existing security mechanisms rather than replacing them. We hope that this first step will stimulate further work aiming at bridging the gap between the two worlds.

\appendices

\section{Proof of Theorem~\ref{thm1}}\label{app_main_proof}

\subsection{Achievability Proof}
The proof is given for a fixed average power $P\leq
\bar{P}$ and transmission rate $R_0$. The key rate is then obtained
by the appropriate maximization. Let $R_s = C_s^{(g)} - \delta$ for some
small $\delta >0$ and $R = R_0 - \epsilon$. We first generate all binary sequences $\{
{\mathbf V} \}$ of length $m R$ and then independently assign each
of them randomly to one of $2^{n R_s}$ groups, according to a
uniform distribution. This ensures that any of the sequences are
equally likely to be within any of the groups. Each secret message
$w \in \{1, \cdots, 2^{n R_s} \}$ is then assigned a group ${\mathbf
V}(w)$. We then generate a Gaussian codebook consisting of $2^{n_1
\left(R_0 - \epsilon \right)}$ codewords, each of length $n_1$
symbols. The codebooks are then revealed to Alice, Bob, and Eve. To
transmit the codeword, Alice first selects a random group ${\mathbf
v}(i)$ of $n_1R$ bits, and then transmits the corresponding
codeword, drawn from the chosen Gaussian codebook. If Alice receives
an ACK bit from Bob, both are going to store this group of bits
and selects another group of bits to send in the next coherence
interval in the same manner. If a NACK was received, this group
of bits is discarded and another is generated in the same manner.
This process is repeated till both Alice and Bob have shared the
same key $w$ corresponding to $nR_s$ bits. We observe that the
channel coding theorem implies the existence of a Gaussian codebook
where the fraction of successfully decoded frames is given by $\frac{m}{n}=\textrm{Pr}\left( R_0 \leq \log_2 \left( 1+h_bP \right) \right),$ as $n_1\rightarrow\infty$. The equivocation rate at the eavesdropper can then be lower bounded as follows.
    \begin{align}
        n R_e   & = H\left(W|Z^n,K^b,G_b^b,G_e^b\right) \no \\
                & \overset{(a)}{=} H\left(W|Z^m,G_b^a,G_e^a\right) \no \\
                & = H\left(W,Z^m|G_b^a,G_e^a\right) - H\left(Z^m|G_b^a,G_e^a\right) \no \\
                & = H\left(W,Z^m,X^m|G_b^a,G_e^a\right) - H\left(Z^m|G_b^a,G_e^a\right) \no \\*
                & \quad - H\left(X^m|W,Z^m,G_b^a,G_e^a\right) \no \\
                & = H\left(X^m|G_b^a,G_e^a\right) + H\left(W,Z^m| X^m,G_b^a,G_e^a\right) \no \\*
                & \quad - H\left(Z^m|G_b^a,G_e^a\right) - H\left(X^m|W,Z^m,G_b^a,G_e^a\right) \no \\
                & \ge H\left(X^m|G_b^a,G_e^a\right) + H\left(Z^m| X^m,G_b^a,G_e^a\right) \no \\*
                & \quad - H\left(Z^m|G_b^a,G_e^a\right) - H\left(X^m|W,Z^m,G_b^a,G_e^a\right) \no \\
                & = H\left(X^m|G_b^a,G_e^a\right) - I\left(Z^m ; X^m|G_b^a,G_e^a\right) \no \\*
                & \quad - H\left(X^m|W,Z^m,G_b^a,G_e^a\right) \no \\
                & = H\left(X^m | Z^m,G_b^a,G_e^a\right) - H\left(X^m|W,Z^m,G_b^a,G_e^a\right) \no \\
                & \overset{(b)}{=} \sum_{j=1}^a H\left(X(j)|Z(j),G_b(j),G_e(j)\right) \no \\*
                & \quad - H\left(X^m|W,Z^m,G_b^a,G_e^a\right) \no \\
                & \overset{(c)}{\ge} \sum_{j \in {\mathcal N}_m} H\left(X(j)|Z(j),G_b(j),G_e(j)\right) \no \\*
                & \quad - H\left(X^m|W,Z^m,G_b^a,G_e^a\right) \no \\
                & = \sum_{j \in {\mathcal N}_m} \big[ H\left(X(j)|G_b(j),G_e(j)\right)  \no \\*
                & \quad - I\left(X(j);Z(j)|G_b(j),G_e(j)\right) \big]  \no \\*
                & \quad - H\left(X^m| W,Z^m,G_b^a,G_e^a\right)  \no \\
                & \ge  \sum_{j \in {\mathcal N}_m} n_1 \big[ R_0 - \log_2 \left( 1 + h_e(j) P \right)
                            - \epsilon \big]  \no \\*
                & \quad - H\left(X^m|W,Z^m,G_b^a,G_e^a\right)  \no \\
                & \ge \sum_{j=1}^{a} n_1 \big\{ \left[ R_0 - \log_2\left( 1 + h_e(j) P \right) \right]^{+}  - \epsilon \big\} \no \\*
                & \quad - H\left(X^m|W,Z^m,G_b^a,G_e^a\right)  \no \\
                & \overset{(d)}{=} n C_s^{(g)} - H\left(X^m|W,Z^m,G_b^a,G_e^a\right) - m \epsilon.\label{lb1}
    \end{align}
In the above derivation, (a) results from the independent choice of
the codeword symbols transmitted in each ARQ frame which does not
allow Eve to benefit from the observations corresponding to the
NACKed frames, (b) follows from the memoryless property of the
channel and the independence of the $X(j)$'s, (c) is obtained by
removing all those terms which correspond to the coherence intervals
$j \notin {\mathcal N}_m$, where ${\mathcal N}_m = \left\{ j \in
\{1, \cdots,a\} : h_b(j) > h_e(j) | \psi = 1\right\}$, where $\psi$ is a binary random variable and $\psi = 1$ indicates that an ACK was received, and (d) follows from
the ergodicity of the channel as $n, m \rightarrow \infty$. Now we
show that the term $H(X^m|W,Z^m,G_b^a,G_e^a)$ vanishes as ${n_1} \to
\infty$ by using a list decoding argument. In this list decoding, at
coherence interval $j$, the wiretapper first constructs a list
${\mathcal L}_j$ such that ${\bf x}(j) \in {\mathcal L}_j$ if $({\bf
x}(i),{\bf z}(i))$ are jointly typical. Let ${\mathcal L}={\mathcal
L}_1 \times{\mathcal L}_2\times\cdots\times{\mathcal L}_a$. Given
$w$, the wiretapper declares that $\hat{{\bf x}}^m=({\bf x}^m)$ was
transmitted, if $\hat{x}^m$ is the only codeword such that
$\hat{{\bf x}}^m \in B(w)\cap {\mathcal L}$, where $B(w)$ is the set
of codewords corresponding to the message $w$. If the wiretapper
finds none or more than one such sequence, then it declares an
error. Hence, there are two types of error events: 1) ${\mathcal
E}_1$: the transmitted codeword ${\bf x}^m_t$ is not in ${\mathcal
L}$, 2) ${\mathcal E}_2$: $\exists {\bf x}^m \neq {\bf x}^m_t$ such
that ${\bf x}^m \in B(w)\cap{\mathcal L}$. Thus the error
probability $\Prob (\hat{{\bf x}}^m \neq {\bf x}^m_t )= \Prob (
{\mathcal E}_1\cup {\mathcal E}_2 ) \leq \Prob ( {\mathcal E}_1) +
\Prob ({\mathcal E}_2)$. Based on the Asymptotic Equipartition
Property (AEP), we know that $\Prob ({\mathcal E}_1) \leq
\epsilon_1$. In order to bound $\Prob ({\mathcal E}_2)$, we first
bound the size of ${\mathcal L}_j$. We let
\begin{align*}
    \phi_j({\bf x}(j)|{\bf z}(j))=\left\{
    \begin{array}{ll}
        1,& \textrm{$({\bf x}(j),{\bf z}(j))$ are jointly typical,} \\*
        0,& \textrm{otherwise.}
    \end{array} \right.
\end{align*}
Now
\begin{align*}
    {\mathbb E}\{\|{\mathcal L}_j\|\}
        & = {\mathbb E}\left\{\sum\limits_{{\bf x}(j)}\phi_j({\bf x}(j)|{\bf z}(j))\right\}\no\\
        & \leq{\mathbb E}\left\{1+\sum\limits_{{\bf x}(j) \neq {\bf x}_t(j)}
                \phi_j({\bf x}(j)|{\bf z}(j))\right\}\no \\
        & \leq 1+\sum\limits_{{\bf x}(j) \neq {\bf x}_t(j)}{\mathbb E}\left\{\phi_j({\bf x}(j)|
        {\bf z}(j))\right\} \no\\
        & \leq 1+2^{{n_1}\left[R_0 - \log_2(1+h_e(j)P)-\epsilon\right]}\no\\
        & \leq 2^{{n_1}\left(\left[R_0-\log_2(1+h_e(j)P) - \epsilon \right]^+ +\frac{1}{n_1} \right)}.
\end{align*}
Hence
\begin{align*}
    {\mathbb E}\{\|{\mathcal L}\|\}
        & = \prod\limits_{j=1}^{a} {\mathbb e}\{\|{\mathcal L}_j\|\} \no \\
        & = 2^{\sum\limits_{j=1}^a n_1\left(\left[R_0-
            \log_2(1+h_E(j)P) - \epsilon \right]^+ + \frac{1}{n_1}\right) }. \no \\
    \Prob ({\mathcal E}_2 )
        & \leq {\mathbb E}\left\{\sum\limits_{x^m \in{\mathcal L}, {\bf x}^m \neq {\bf x}^m_t}
            \Prob ({\bf x}^m \in B(w)) \right\} \no \\
        & \overset{(a)}\leq {\mathbb E}\left\{\|{\mathcal L}\|2^{-n R_s}\right\} \no \\
        & \leq 2^{-n R_s}2^{\sum\limits_{j=1}^a n_1\left(\left[R_0-\log_2(1+h_e(j)P) - \epsilon \right]^+ +
            \frac{1}{n_1} \right) }\no\\
        & \leq 2^{-n \left(R_s -\frac{1}{c}\sum\limits_{j=1}^a \left(\left[R_0
            -\log_2(1+h_e(j)P) - \epsilon \right]^+ + \frac{1}{n_1} \right) \right)} \no \\
        & = 2^{-n \left(R_s -\frac{1}{c}\sum\limits_{j=1}^a \left(\left[R_0 -\log_2(1+h_e(j)P) \right]^+ + \frac{1}{n_1} \right) + \frac{|{\mathcal N}_m| \epsilon}{c} \right)},
\end{align*}
\noindent where (a) follows from the uniform distribution of the codewords in
$B(w)$. Now as $n_1 \to \infty$ and $a \to \infty$, we get \[ \Prob
({\mathcal E}_2 ) ~\le ~ 2^{-n \left(C_s^{(g)} - \delta - C_s^{(g)} + a \epsilon
\right)} ~=~ 2^{-n(c \epsilon -\delta)}, \] where $c = \Prob ( h_b >
h_e)$. Thus, by choosing $\epsilon > (\delta / c)$, the error
probability $\Prob ({\mathcal E}_2 ) \to 0$ as $n \to \infty$. Now
using Fano's inequality, we get $H(X^m|W,Z^m,G_b^a,G_e^a) ~\leq~ n\delta_{n} \to 0$ as $m,n \to \infty$. Combining this with (\ref{lb1}), we get the desired result.

\subsection{Converse Proof}

We now prove the converse part by showing that for any perfect secrecy rate $R_s$ with equivocation rate $R_e > R_s - \epsilon$ as $n,m \rightarrow \infty$, there exists a transmission rate $R_0$, such that
\begin{equation*}
    \begin{split}
        R_s & ~\leq~ \mathbb{E}\Big\{\left[R_0 - \log_2\left(1+h_eP\right)\right]^{+} \no \\
            & \qquad \mathbb{I}\left(R_0 \leq \log_2\left(1+h_bP\right) \right)\Big\}.
    \end{split}
\end{equation*}

Consider any sequence of $(2^{nR_s},m)$ codes with perfect secrecy rate $R_s$ and equivocation rate $R_e$, such that $R_e > R_s - \epsilon$ as $n \rightarrow \infty$.
We note that the equivocation $H(W|Z^n,K^n,G_b^b,G_e^b)$ only depends on the marginal distribution of $Z^n$, and thus does not depend on whether $Z(i)$ is a physically or stochastically degraded version of $Y(i)$ or vice versa. Hence we assume in the following derivation that for any fading state, either $Z(i)$ is a physically degraded version of $Y(i)$ or vice versa (since the noise processes are Gaussian). Thus we have
\begin{align*}
    nR_e    & = H(W|Z^b,K^n,G_b^b,G_e^b) \no \\
            & \overset{(a)}= H(W|Z^m,G_b^a,G_e^a) \no \\
            & \overset{(b)}{\le} H(W|Z^m,G_b^a,G_e^a) - H(W|Z^m,Y^m,G_b^a,G_e^a) \no \\
            & \quad + m\delta_m \no \\
            & = I(W;Y^m|Z^m,G_b^a,G_e^a) + m\delta_n  \no \\
            & \overset{(c)}{\le} I(X^m;Y^m|Z^m,G_b^a,G_e^a) + m\delta_m  \no \\
            & = H(Y^m|Z^m,G_b^a,G_e^a)  \no \\
            & \quad - H(Y^m|X^m,Z^m,G_b^a,G_e^a) + m\delta_m  \no \\
            & = \sum_{i=1}^a [ H(Y(i)|Y^{i-1},Z^m,G_b^a,G_e^a)  \no \\
            & \quad - H(Y(i)|Y^{i-1},X^m,Z^m,G_b^a,G_e^a) ] + m\delta_m  \no \\
            & \overset{(d)}{\le} \sum_{i=1}^a [ H(Y(i)|Z(i),G_b(i),G_e(i) \no \\
            & \quad - H(Y(i)|X(i),Z(i),G_b(i),G_e(i)) ] + m\delta_m  \no \\
            & = \sum_{i=1}^a I(X(i);Y(i)|Z(i),G_b(i),G_e(i)) + m\delta_m  \no \\
            & \overset{(e)}{=} \sum_{i=1}^a I(X(i);Y(i)|G_b(i),G_e(i))  \no \\
            & \quad - I(X(i);Z(i)|G_b(i),G_e(i)) + m\delta_m  \no \\
            & {\le}  \sum_{i=1}^a R_0 - \log_2(1+h_e(i)P) + m\delta_m  \no \\
            & {\le}  \sum_{i=1}^a [ R_0 - \log_2(1+h_e(i)P) ]^+ + m\delta_m  \no \\
    R_e     &\overset{(f)}{\le} \mathbb{E}\big\{\left[R_0 - \log_2\left(1+h_eP\right)\right]^{+} \no \\
            & \quad \mathbb{I}\left(R_0 \leq \log_2\left(1+h_bP\right) \right)\big\} + \beta\delta_m,
\end{align*}
\noindent where $\beta = \textrm{Pr}(R_0 \leq \log_2(1+h_bP))$. In the above derivation, (a) results from the independent choice of the codeword symbols transmitted in each ARQ frame which does not allow Eve to benefit from the observations corresponding to the NACKed frames, (b) follows from Fano's inequality, (c) follows from the data processing inequality since $W \to X^m \to (Y^m,Z^m)$ forms a Markov chain, (d) follows from the fact that conditioning reduces entropy and from the memoryless property of the channel, (e) follows from the fact that $I(X;Y|Z) = I(X;Y) - I(X;Z)$ as shown in~\cite{Wyner1975}, (f) follows from ergodicity of the channel as $m,n \rightarrow \infty$. The claim is thus proved.

\section{Proof of Decorrelation}\label{app_decor_proof}

In this appendix, we show that employing multiple transmit antennas makes the correlation between Eve's and Bob's channel power gains converge to zero, in a mean-square sense, as the number of antennas $N$ goes to $\infty$. Let $l_1 = |g_b^{eq}|^2$ and $l_2 = |g_e^{eq}|^2$. Assuming all $\theta$'s to be uniformly distributed in the interval $[-\pi,\pi]$, we get,
\begin{align}
    l_1 & = \frac{1}{N}\left[\left\vert\sum\limits_{i=1}^{N}\cos\left(\theta_{iR}+\theta_{iB}\right)\right\vert^2
            + \left\vert\sum\limits_{i=1}^{N}\sin\left(\theta_{iR}+\theta_{iB}\right)\right\vert^2\right] \no \\
        & = \frac{1}{N}\Bigg[N +
            2\sum\limits_{i=1}^{N-1}\sum\limits_{j=i+1}^{N}\Big\{\cos\left(\theta_{iR}+\theta_{iB}\right)
            \cos\left(\theta_{jR}+\theta_{jB}\right) \no \\*
        & \qquad \qquad +
            \sin\left(\theta_{iR}+\theta_{iB}\right)\sin\left(\theta_{jR}+\theta_{jB}\right)\Big\}\Bigg] \no \\
        & = 1 + \frac{2}{N}
            \sum\limits_{i=1}^{N-1}\sum\limits_{j=i+1}^{N}
            \cos\left(\theta_{iR}+\theta_{iB}-\theta_{jR}-\theta_{jB}\right). \label{l1eq}
\end{align}
\noindent Similarly for $l_2$,
\begin{align}
    l_2 & = 1 + \frac{2}{N}
            \sum\limits_{i=1}^{N-1}\sum\limits_{j=i+1}^{N}
            \cos\left(\theta_{iR}+\theta_{iE}-\theta_{jR}-\theta_{jE}\right).
    \label{l2eq}
\end{align}
Now, taking the expectation of~(\ref{l1eq}) and~(\ref{l2eq}) with respect to the random phases applied on the transmit antenna array $\theta_{iR}$ for given values of $\theta_{iE}$'s and $\theta_{iB}$'s, we get $\mathbb{E}\left(l_1\right) = \mathbb{E}\left(l_2\right) = 1,$ and
\begin{align*}
    \mathbb{E}\left(l_1\right)
        & = \mathbb{E}\left(l_2\right) = 1, \\
    \mathbb{E}\left(l_1l_2\right)
        & = 1 + \frac{2}{N^2}
            \sum\limits_{i=1}^{N-1}\sum\limits_{j=i+1}^{N}\cos\left[\left(\theta_{iB}-\theta_{iE}\right) - \left(\theta_{jB}-\theta_{jE}\right)\right], \\
    \mathbb{E}\left(l_1^2\right)
        & = \mathbb{E}\left(l_2^2\right) = 1 + \frac{2}{N^2}\frac{N(N-1)}{2} = 1 + \frac{N-1}{N}.
\end{align*}
So, the variance of $l_1$ and $l_2$ is given by
\begin{equation*}
    \textrm{var}\left(l_1\right) = \textrm{var}\left(l_2\right) = \sigma^2_{l_1} = \sigma^2_{l_2} = \frac{N-1}{N}.
\end{equation*}
Therefore, the correlation coefficient $\rho$ between the channels' power gains is given by
\begin{align*}
    \rho    & = \frac{\mathbb{E}\left(l_1l_2\right) -
                \mathbb{E}\left(l_1\right)\mathbb{E}\left(l_2\right)}
                {\sqrt{var\left(l_1\right)}\sqrt{Var\left(l_2\right)}}\no\\
            & = \frac{2}{N(N-1)}
                \sum\limits_{i=1}^{N-1}\sum\limits_{j=i+1}^{N}\cos\left[\left(\theta_{iB}-\theta_{iE}\right) - \left(\theta_{jB}-\theta_{jE}\right)\right]\no\\
            & = \frac{2}{N(N-1)} \sum\limits_{i=1}^{N-1}\sum\limits_{j=i+1}^{N}\cos\left[\Delta_i - \Delta_j\right],
\end{align*}
\noindent where $\Delta_i = \theta_{iB}-\theta_{iE}$ and $\Delta_j = \theta_{jB}-\theta_{jE}$. Assuming $\theta_{iB}, \theta_{iE}, \theta_{jB}, \theta_{jE}$ are all independent, and uniformly distributed in the interval $[-\pi,\pi]$, and taking the expectation of $\rho$ over them, we get
\begin{equation}
    \mathbb{E}\left(\rho\right) = 0.
\end{equation}
The divergence of $\rho$ around its mean is given by
\begin{align}
    \textrm{var}(\rho)
        & = \sigma^2 \no\\
        & = \frac{4}{N^2(N-1)^2}\sum\limits_{i=1}^{N-1}\sum\limits_{j=i+1}^{N}\textrm{var}\left(\cos\left(\Delta_i
            - \Delta_j\right)\right)\no\\
        & = \frac{4}{N^2(N-1)^2}.\frac{N(N-1)}{2}.\frac{1}{2}\no\\
        & = \frac{1}{N(N-1)}.\label{corr_conv}
\end{align}
\noindent Thus, the standard deviation of $\rho$ is given by $ \sigma = \frac{1}{\sqrt{N(N-1)}} \simeq \frac{1}{N}$. It is evident from~(\ref{corr_conv}) that $\textrm{var}(\rho)$ goes to zero as $N \rightarrow \infty$. That is, the correlation coefficient $\rho$ converges, in a mean-square sense, to zero.


%
%

\ifCLASSOPTIONcaptionsoff
  \newpage
\fi




\bibliographystyle{IEEEtran}
\bibliography{refs}

\end{document}